\documentclass{article}

\usepackage{arxiv}
\usepackage{epigraph}
\usepackage[utf8]{inputenc} 
\usepackage[T1]{fontenc}    
\usepackage{hyperref}       
\usepackage{url}            
\usepackage{booktabs}       
\usepackage{amsfonts}       
\usepackage{nicefrac}       
\usepackage{microtype}      
\usepackage{graphicx}
\usepackage{doi}
\usepackage{xspace}
\usepackage{subcaption}
\usepackage{pgfplots}
\usepackage{pgfplotstable}
\usepackage{xcolor}
\usepackage{booktabs}
\pgfplotsset{compat=1.9}
\usetikzlibrary{patterns}
\usepackage{amsmath}
\usepackage{authblk} 
\usetikzlibrary{shapes.geometric}
\usetikzlibrary{arrows.meta,arrows}

\title{Tit-for-Token: Understanding Fairness when Forwarding Data by Incentivized Peers in Decentralized Storage Networks}

\author[1]{Vahid Heidaripour Lakhani}
\author[2]{Arman Babaei}
\author[1]{Leander Jehl}
\author[3]{Georgy Ishmaev}
\author[2]{Vero Estrada-Galiñanes}

\affil[1]{University of Stavanger, Norway}
\affil[2]{École polytechnique fédérale de Lausanne(EPFL), Switzerland}
\affil[3]{Delft University of Technology, Netherlands}

\renewcommand{\shorttitle}{Tit-for-Token}

\hypersetup{
pdftitle={TIT-FOR-TOKEN: UNDERSTANDING FAIRNESS WHEN FORWARDING DATA BY INCENTIVIZED PEERS IN DECENTRALIZED STORAGE NETWORKS},
pdfsubject={},
pdfauthor={Vahid Heidaripour Lakhani, Arman Babaei, Leander Jehl, Georgy Ishmaev, Vero Estrada-Galiñanes},
pdfkeywords={Fairness, Bandwidth Incentives, Token-based Incentives, Networked Economy, Web3 Incentives, Reciprocity, Tit-for-Tat, Monetary-based Incentives, Decentralized storage networks, Prefix-based Routing Networks, Libp2p},
}

\begin{document}

\def\mp{\ensuremath{\omega}\xspace}
\def\rewardfairness{Reward-Fairness\xspace}
\def\incomefairness{Income-Fairness\xspace}
\newtoggle{comments}
\iftoggle{comments}{
\long\def\todo#1 { {\bf TODO:} [{\color{red} #1}] }
\long\def\lj#1{ {\bf LJ: } [{\color{olive} \em #1}]}
\long\def\veg#1{ {\bf VEG: } [{\color{blue} \em #1}]}
\long\def\gi#1{ {\bf GI: } [{\color{teal} \em #1}]}
\long\def\vhl#1{ {\bf VHL: } [{\color{orange} \em #1}]}
\long\def\ab#1{ {\bf AB: } [{\color{magenta} \em #1}]}
}{
\long\def\todo#1{}
\long\def\lj#1{}
\long\def\veg#1{}
\long\def\gi#1{}
\long\def\vhl#1{}
\long\def\ab#1{}
}

\newcommand{\commonBits}{\texttt{commonBits}}

\newcommand{\tft}{\textit{Tit4Tat}}
\newcommand{\tftok}{\textit{Tit4Tok}}

\newcounter{sipcnt}
\newenvironment{sip}[1][]{\refstepcounter{sipcnt}
\par\noindent\textbf{Swarm improvement proposal:}
\textit{#1}}{}

\newcommand{\mpinsection}{\texorpdfstring{$\omega$}{Omega}}

\newcounter{obscnt}
\newenvironment{observation}[1][]{\refstepcounter{obscnt}\par\noindent\textbf{Observation~\theobscnt:} \textit{#1}}{}

\maketitle

\begin{abstract}
\noindent
Decentralized storage networks offer services with intriguing possibilities to reduce inequalities in an extremely centralized market. The challenge is to conceive incentives that are fair in regard to the income distribution among peers. Despite many systems using tokens to incentivize forwarding data, like Swarm, little is known about the interplay between incentives, storage-, and network-parameters. This paper aims to help fill this gap by developing Tit-for-Token (Tit4Tok), a framework to understand fairness. Tit4Tok realizes a triad of altruism (acts of kindness such as debt forgiveness), reciprocity (Tit-for-Tat's mirroring cooperation), and monetary rewards as desired in the free market. Tit4Tok sheds light on incentives across the accounting and settlement layers. We present a comprehensive exploration of different factors when incentivized peers share bandwidth in a libp2p-based network, including uneven distributions emerging when gateways provide data to users outside the network. We quantified the \incomefairness with the Gini coefficient, using multiple model instantiations and diverse approaches for debt cancellation. We propose regular changes to the gateway neighborhood and show that our shuffling method improves the \incomefairness from 0.66 to 0.16. We quantified the non-negligible cost of tolerating free-riding (altruism). The performance is evaluated by extensive computer simulations and using an IPFS workload to study the effects of caching.
\end{abstract}


\section{Introduction}
\label{sec:new-introduction}

\epigraph{If your friend lent you money in your distress, ought you to lend him money in his? How much ought you to lend him? When ought you to lend him? Now, or tomorrow, or next month? And for how long a time?}{Adam Smith, The theory of moral sentiments}

\noindent Open decentralized systems (ODS)\footnote{We consider open decentralized systems (ODS) as a broad label for distributed systems where there is no centralized control over system components and functions, and all participants can join and leave the system easily. These include p2p systems, permissionless blockchain protocols, and other decentralized solutions.} such as the Interplanetary File System (IPFS) or the Swarm network propose a tantalizing vision of a decentralized web and a fair data economy through large-scale collaborative ecosystems that depend on incentivized users, content generators, and operators of data-sharing platforms facilitated by peers moving and storing data across a network~\cite{ipfs_and_friends,tron2021swarm,benett2014ipfs,jaiman2022user}. 
Furthermore, evidence shows that despite some problems with the reliability and manageability of autonomous peer operators forming peer-to-peer (p2p) networks, these networks provide very cost-effective solutions. 
This is evident from the observation that even large companies benefit from these networks, especially in edge-computing, content distribution networks, and other systems using mechanisms similar to those used in the BitTorrent network~\cite{broberg2009metacdn,eshel2010panache,garcia2015edge,understanding_hybrid_cdn-p2p,yadgar2019modeling}.

Many ODS, including the systems mentioned above, have a networking stack that depends on the libp2p project~\cite{noauthor_libp2p_nodate}. 
This modular library provides key components to build decentralized networks equally accessible from anywhere in the world.
On top of the networking layer, many systems include some kind of incentivization layer.
For instance, Swarm uses the SWAP protocol and the postage stamps to incentivize bandwidth and storage sharing respectively. 
The topic of incentives has received a lot of attention in systems research in part thanks to the advancements in blockchain protocols and token-based incentivization mechanisms~\cite{zhang2009review,rahman2010improving,bano2019sok, compounding_wealth_POS_Giulia_Fanti, ihle2023incentive}.
Nonetheless, the literature is vague when it comes to discerning how the resources shared in the network are financed and if peer operators receive a fair reward for contributing to the ecosystem.
We know that imbalances in incentives can cause centralization problems, e.g., consensus power concentration, routing centralization, wealth concentration, bandwidth concentration, etc.  
However, a taxonomy of centralization in public systems~\cite {SAI2021102584} showed that consensus power has been widely studied, leaving, in comparison, a research gap for the factors that affect bandwidth incentives.  

This paper aims to help fill this gap by developing \emph{Tit-for-Token} or, in short, \tftok, a framework to understand fairness when forwarding data by incentivized peers in decentralized storage networks. 
We also present the results from a comprehensive exploration of a compound of incentive mechanisms found in real-world networks using realistic workloads. 

Incentives often play a significant role in motivating peer operators to participate in a network.
Well-designed incentives can be used to reinforce motivation.  
We define the \emph{triad of altruism, reciprocity, and free enterprise} as required incentives for a more fair data economy. 
In other words, our framework realizes a triad that comprises acts of kindness such as debt forgiveness, mirroring cooperation such as in standard Tit-for-Tat incentives, and monetary rewards as desired in the free market.  
We think that realizing this triad could potentially bring closer the vision of a fair data economy found in the Swarm network community.
While the ideas of a fair data economy, data sovereignty, and decolonization of the digital space are flourishing among society, e.g., software developers, content creators, artists, investors, and EU policymakers, the literature lags behind.
This is unfortunate for advancing networked systems research with societal impact.
The majority of the papers on incentives published during the last three decades only focus on one or at most two aspects of the triad~\cite{ihle2023incentive}. 
For an illustrative example, we need to retrace incentives to the reciprocity of the Tit-for-Tat (from here on abbreviated \tft) found in BitTorrent~\cite{cohen2003incentives}.
Despite the success of BitTorrent, and the good arguments about \tft{}, this mechanism, or many others that have been proposed later, is not solid enough to provide an alternative to the asymmetric wealth distribution of the current data economy. 

By paraphrasing Adam Smith, 
\emph{if a peer served your network request in your distress, ought you to serve his request in his? How much ought you to serve him? When ought you to serve him? Now, or tomorrow, or next month? And for how long?}
These questions are closely related to the rationale behind \tftok{}.
Altruism brings up questions like: Ought you to forgive someone's debt? 
Will forgiveness create imbalances?
How often do you ought to tolerate free riders? 
Reciprocity, instead, raises questions like: Ought you to remember the favors somebody made for you?
Ought you to payback with another favor?
Finally, free enterprise triggers the warning:
Debts, or favors, are understood better in monetary terms.
Ultimately, ought you to serve two peer requests equally?

In more detail, our contributions are:
\vspace{0.5ex}
\noindent $\bullet$ \emph{Incentive model:} We introduce an abstract model to incentivize network peers to realize the triad of altruism, reciprocity, and free enterprise.
\noindent $\bullet$ \emph{Incentive toolkit:} We operationalize the Tit4Tok concept with a comprehensible open-sourced toolkit to simulate different instantiations of the model and emulate real decentralized network workloads.
\noindent $\bullet$ \emph{Fairness analysis:} We investigate potential sources for uneven distributions and their effect on fairness like for example, the gateways neighborhood. One of the sources of uneven distributions is when a few peers are originating large amounts of requests. This situation could happen when peers act as a gateway to take requests from clients that do not participate in the network and access a gateway, for example via a normal browser.
\noindent $\bullet$ \emph{Measurements:} We measure the Gini coefficient to quantify the income fairness in the accounting and settlement layer to 1) study the interplay of  the income received by peers with network parameters, and 2) compare the distribution of received tokens with and without reciprocity and free service. An \incomefairness of 0 reflects perfect fairness while 1 reflects maximal inequality.
\noindent $\bullet$ \emph{Shuffling:} We proposed shuffling to regularly change the gateway neighborhood and show that it can improve \incomefairness from 0.66 to 0.16.
\noindent $\bullet$ \emph{Altruism:} We show that the cost of providing a limited free service may distributed unevenly among peers, possibly worsening income-fairness from 0.34 to 0.63, and propose pairwise limits based on address distance that distributes the cost more fairly (0.47).
\noindent $\bullet$ \emph{Gateway Cliques:} We investigate the benefit and centralization risk connected gateways pose, compared to the benefits powerful operators can gain from coordinating multiple gateways.
\noindent $\bullet$ \emph{Caching:} We show that caching can smoothen imbalance due to peer distribution in the address space, but increases imbalance due to gateways.
\vspace{0.5ex}

We presented preliminary results on fairness in Swarm at~\cite{workshop_fairness_ours}, but all of the above contributions are new or significantly extended in this paper.


\section{Tit-for-Tat: Preliminaries and Prior Work}
\gi{We mention 'ODS' in the introduction but never use it after, maybe it makes sense to mention it in this section and in the conclusion.}
Incentive mechanisms are a way to coordinate participation, increase cooperation, and limit the selfish behaviors of network participants. 
This section gives a broad overview of \tft{} and the general challenges and open problems of sharing computational resources in decentralized storage networks. 
Finally, we summarize the literature gaps and discuss what motivates \tftok{}, particularly in the scope of bandwidth incentives. 

Peer-to-peer (p2p) networks are computationally-based human social systems that create a common pool of resources available for community participants in an 
open membership and, commonly, a permissionless environment.
An unmanaged commons, where participants do not control the overuse of resources, risks the extremely unfair event of depletion of resources, resulting in a network collapse known as the ``tragedy of the commons''~\cite{hardin1998extensions} (a discussion is found in Appendix~\ref{app:Hardin}).
Thus, it is crucial to manage the computational resources of open and permissionless networks via effective incentive mechanisms. 

\subsection{Tit-for-Tat: Strength and Limitations}
The \tft{} 
mechanism has been intensively used in p2p networks, e.g., BitTorrent~\cite{cohen2003incentives}, with the general belief that it efficiently discourages free-riding, i.e., peers who only consume resources without giving back.
It has a simple strategy in which each participant first cooperates and then mirrors, or reciprocates, the immediately observed behavior from its interacting peers, likely to incentivize mutual cooperation. 
\tft{} works by punishing bad behaviors in the future, i.e., \emph{cheat me first, and I will cheat you back}. 
This concept, referred to as the ``shadow of the future,'' was largely studied by Axelrod in computer tournaments playing the Prisoner's Dilemma cooperation game ~\cite{axelrod1981evolution}. 
Later, the Pavlov strategy proposed a more robust strategy that included some degree of forgiveness or generosity between peers ~\cite{nowak1993strategy}. 

\paragraph{Altruism vs Free-Riding.\,}
The cooperation that surges from \tft{} is known as ``reciprocally altruistic behavior''~\cite{trivers1971evolution}.
But does it work in practice? 
One of the main criticisms is that \tft{} can be easily subverted by participants who change their client's code to cheat (fail to reciprocate) their peers. 
The \tft{} in BitTorrent can induce free riding~\cite{jun2005incentives}, and entire files can be downloaded without reciprocating in a cheap free-riding attack~\cite{bitThief,piatek2007incentives-bitTyrant}.   
The problem of selfish and misbehaving nodes was widely studied in the literature, which offers a plethora of strategies, including punishments and/or variations of the \tft{} to mitigate misbehaviors~\cite{jun2005incentives,zghaibeh2008revisiting,amortized_tit-for-tat, enhancing_tit-for-tat_free-riding, o-torrent, rescuing_tit-for-tat_source_coding, enhancing_tit-for-tat_in_bittorrent, feldman2005overcoming}.
On the contrary, altruistic behavior provides an alternative narrative, which explains why networks do not collapse~\cite{vassilakis2007modelling}.
Moreover, a generalized reciprocity behavior, in which peers do favors without direct expectations while relying only on somebody else willing to do a favor to them, can explain why peers can tolerate some free-riding behavior~\cite{jian2008}.
Tribler presented down-to-earth expectations about altruism with its social group incentives based on the ``kinship fosters cooperation'' argument~\cite{trib2008}.  
Scientists found that even a small amount of altruism effectively improved the performance of a p2p live streaming service~\cite{chu2004considering}. 
As other scientists noted, substantial research has been dedicated to discouraging selfish behavior with complex technical solutions or disregarding the cost overhead \cite{oliveira2013can}.
Considering all the above, our model \tftok{}, which builds on \tft{} and real-world decentralized networks, motivates altruistic behaviors continuously, and tolerates some free-riding.

\paragraph{Mechanism Dependencies on Upper Layers.\,} 
\tft{} reciprocates the immediately observed behavior, and going beyond that requires a public history of behaviors, a robust identification layer, or other complex mechanisms often found in reputation-based incentives. 
The overarching question is which peer is trustworthy or at least offers the best cost-quality service relationship~\cite{stokkink2020foundations}. 
Reputation-based systems present multiple challenges and can harm participants.
For example, "Sybil attacks," in which a single entity controls multiple fake identities either to inflate its reputation value or discredit other participants\cite{prunster2022total}.
Despite the centralized control, even YouTube cannot stop abusers from illicit monetization exploits like selling accounts\cite{chu2022behind}.
Our paper focuses on the networking and incentive layers without adding dependencies to more complex mechanisms. 
\tftok{} has a mutual accounting layer but does not depend on upper layers like other reputation-based incentives. 
The mutual accounting layer could be further improved with ideas such as indirect reciprocity~\cite{nasrulin2023sustainable} to address manipulation by Sybils without the cost of a global reputation layer.


\paragraph{Novel Business Models and Stakeholders.\,}
\tft{} drives peers in BitTorrent to exchange content of mutual interest. 
Even if reciprocating bandwidth instead of content is an improvement~\cite{amortized_tit-for-tat}, the model behind it is a restricted version of a barter economy that suffers from the double coincidence of wants, i.e., peers would reward a forwarded chunk with another forwarded chunk. 
Thus, \tft{} impairs complex business developments. 
What if a node sharing bandwidth is not interested in bandwidth for itself at that particular time or if the node participates in the network mostly to generate wealth (income)? 
Monetary or credit-based incentives can address the questions above by enabling more practical transactions among peers than in bartering economies. 
This area has generated significant interest~\cite{sirivianos2009robust,ghosh2014torpath,kopp2016koppercoin, miller2014permacoin}, especially with recent research focusing on cryptocurrencies and token-based incentives.  
Our paper acknowledges that 
participants may take on different roles, especially in large-scale systems; for example, some peers prefer to optimize bandwidth provision while others provide storage provision without forgetting those interested in consuming resources. 
Thus, we evaluate, for example, if peers acting as gateways change the rules of the game by bringing centralization or income inequalities.

\subsection{Literature Gap} 
A recent ACM Computing Survey, which reviewed p2p incentive mechanisms based on monetary, reputation, and service published between 1993 and 2022, reported a large number of papers focusing on \tft{} to improve service quality, many papers covering auction-based monetary incentives, and close to zero papers focusing on the interplay between DHTs and incentives~\cite{ihle2023incentive}.     
Further, the results found in the literature are difficult to reproduce due to the lack of unavailable or discontinued simulation tools. 
To our disappointment, free services and societal aspects are largely ignored, e.g., the term "free," which appeared more than 60 times, is only used to discuss free-riding mitigation schemes. 

We argue that despite the abundant literature about \tft{}, researchers have not investigated deeply how a triad of altruism, reciprocity, and free enterprise can co-exist in a well-rounded incentive mechanism for decentralized networks. 
Hence, our paper differs in a number of fundamental ways from previous research on incentives. 
We extend or generalize reciprocal behavior to tolerate some free-riding for a limited time. 
This is comparable with freemium services, in which the initial service is free of charge, but a premium is charged for additional services. 
In this paper, peers may accept to move a few chunks per connection for free during a short time period, and requests done after reaching the thresholds are not free.
By relaxing security concerns about free-riding and facilitating a limited free service without judging the reasons of the request, our incentives are more aligned with societal issues and with common strategies to increasing adoption. 
First, we believe that networks that provide some degree of free service contribute to secure the rights of freedom of expression and universal access. 
Second, the development of an inclusive society needs technologies that can be used even by the most marginalized.
Third, empirical research in the mobile app market showed that the freemium option is a cost-effective business model to increase sales without incurring significant marketing costs~\cite{liu2014effects}.  
Thus, limited free services may attract consumption and service adoption, helping achieve desirable network effects.
Fourth, previous experiences in Gnutella showed that most users were free riders and did not adopt anti-free-riding measurements despite recommendations~\cite{hughes2005free}.



\section{Decentralized Storage Networks}
\label{sec:decentralized_storage}

This section introduces the terminology used in this paper, gives an overview of content routing in the context of decentralized storage networks, and discusses the incentives in Swarm. 

\paragraph{Terminology.\,} A \emph{node} (or its equivalent \emph{peer}) connects to other peers (\emph{neighbors}) in a peer-to-peer network. The network depends on participants sharing bandwidth to move data efficiently through the network for serving requests between data consumers (\emph{user}) and suppliers (\emph{storers} or storage nodes). 
Users may become \emph{peer operators} themselves by installing a \emph{network client} to have their own peer (\emph{originator}) connect directly to the network and originate  requests.
Alternatively, users may choose not to run their own peer and use a third party \emph{gateway}.
Gateways serve as an interface that connects the decentralized network to the outside world, including clients, applications, and other networks.
Therefore, gateways receive requests from the outside and originate requests on behalf of users.  
Any piece of content (\emph{chunk}) that is in the local store of a node becomes available to other nodes after being push-synced to the network. 
A storer node is responsible for storing a chunk if the chunk address falls within the node’s area of responsibility.
Unless there is an end-to-end connection between originators and storer, the chunk will travel one or more \emph{hops} following a \emph{routing path}.
The peers in the routing path (\emph{forwarders}) perform \emph{forwarding actions}) to forward the chunk one hop from its current location on the path to the next closer location until it reaches the destination. 
Such content-delivery service benefits the community of users, content creators, and other stakeholders; ultimately, it enables the network storage service that pays to the storer. 
Bandwidth, therefore, is a valuable resource for decentralized storage networks.

\subsection{Peer-to-Peer Networking Stack}
The open-sourced libp2p library~\cite{noauthor_libp2p_nodate} maintained by Protocol Labs includes many components to implement the network layer of any decentralized storage system. 
This resource is used by many systems, including the InterPlanetary File System and Swarm to do the peer and content discovery and routing.
One key component is the Kad-DHT, which implements the Kademlia Distributed Hash Table (DHT) subsystem largely based on the Kademlia~\cite{maymounkov2002kademlia}, and expanded with notions from S/Kademlia~\cite{baumgart2007s}.
The library manage connections at the transport layer, tolerating multiple protocols and protecting peers from some security problems.

The Kademlia distributed hash table (DHT) allows to reach any other network peer in a few hops while maintaining a small number of direct connections. 
Each peer $v$ has a routing table that contains the addresses of other peers but without requiring maintaining a direct connection with them. 
The routing table is organized based on a prefix length and a distance metric and maintained in a $k$-bucket data structure.
The $b$-bit address space is organized with $k$-buckets, where a bucket contains up to $k$ peers' addresses that share a prefix of length $i$ with peer $v$ address, with $i \in [0,b]$.
The \emph{distance metric} helps to find the closest peers to a specific key in the routing table. 
The distance between two keys is the bitwise exclusive-or (XOR) of the SHA-256 hash of the two keys, for example a distance 0 means that both keys are identical, and a distance 1 means that one bit is different.
Peers close to each other in the address space form a \emph{neighborhood} that is responsible of storing multiple chunk replicas. 
Appendix~\ref{app:A_routing_table} provides insights about the routing tables in Kademlia and their construction process.

\subsection{Swarm and its SWAP Protocol}

Swarm is a peer-to-peer network with ongoing efforts to provide storage and communication services globally. 
The mission of the project, as published on its website~\cite{tron2021swarm}, is: ``to shape the future towards a self-sovereign global society and permissionless open markets by providing scalable base-layer data storage infrastructure for the decentralized internet.'' 
According to the writings of the Swarm founder~\cite{bookofswarm}, Swarm aims at developing a fair data economy, defined as ``an economy of processing data characterized by fair compensation
of all parties involved in its creation or enrichment.''

\paragraph{Networking Layer.\,} Swarm was designed to thoroughly integrate with the devp2p Ethereum network layer as well as with the Ethereum blockchain for domain name resolution (using ENS).
Later, it adopted the libp2p library with some features relevant to this paper that distinguish Swarm from other networks summarized here. 
First, the network distributes the content through incentivized peers, which sync chunks directly in the DHT up to the redundantly synced area of responsibility for each chunk. 
This allows content-addressing with fine-granularity, that in term can reduce imbalances caused by popular files, since the content is distributed among all the peers responsible for the chunks.
Second, chunks are synced and retrieved with the aid of incentivized peers, which forward chunks using a variant of Kademlia called forwarding Kademlia. 
An originator initiates a request that is relayed via forwarding nodes $F_0, ..., F_n$ all the way to storer node S, the storer node closest to the chunk address. 
The chunk is then delivered by being passed back along the same route to the downloader.
It is claimed that forwarding Kademlia provides some degree of ambiguity and deniability to any peer sending a chunk.

\paragraph{Incentives Layer.\,} Swarm includes a built-in incentives layer, which includes bandwidth incentives to achieve speedy and reliable data provision and storage incentives to ensure long-term data preservation. 
Since Swarm's initial attempt to incentivize forwarding and storing chunks in the Swap, Swear, and Swindle protocol~\cite{tron2016swap}, various consistent software development iterations have followed, combining off-chain communication and settlements with on-chain registration and enforcement. 
Swarm uses BZZ, an ERC-20 token created on the Ethereum network that can be bridged to other networks through Omnichain for Gnosis Chain or Celer Network for Binance Smart Chain. 
The supply of BZZ is defined by bonding curves that may decrease or increase the token supply. 
The incentive system is enforced through smart contracts on the Gnosis Chain blockchain and powered by the xBZZ token~\footnote{xBZZ is BZZ bridged to the Gnosis Chain using OmniBridge.}.
In order to upload and store data to Swarm, a user needs to have postage stamps. 
Postage stamps can be purchased in batches with xBZZ. 
The monetary incentives for storing data are auction-based and are not considered in this paper. 

\paragraph{Business Layer.\,} The Swarm Foundation aims to enable a scalable and self-sustaining open-source infrastructure for a supply-chain economy of data that provides end-users with a speedy and safe service.
Its business model opens an intriguing possibility: developers get tools for creating and hosting decentralized applications (dApps), NFT metadata, and media files with zero hosting cost, independently of the number of people accessing a dApp, i.e., one person or one million would cost the same, due to in-built crypto-economic incentives of the network. 
The incentivized infrastructure encourages decentralization, inclusivity, and privacy as it relieves developers from the need to seek data monetization models and the support from the exclusive and concentrated venture capital to develop dApps or share content. 




\section{Tit-for-Token}
\label{sec:tit-for-token-model}

This section presents the design of \emph{\tftok{}}, its rationale, and the system model. 

\subsection{Design Overview}
\begin{figure*}
    \centering    \includegraphics[width=0.95\textwidth]{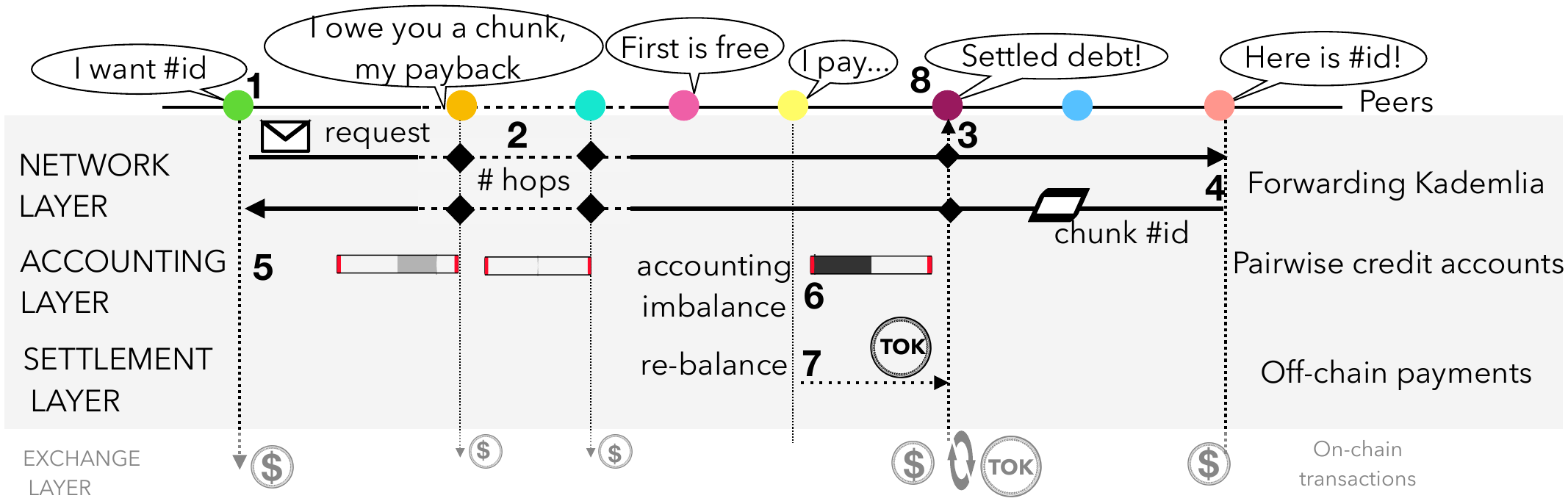}

    \caption{\tftok{} includes  routing and accounting layers. The settlement layer transfers tokens through off-chain payments. The exchange layer is excluded. 1) A gateway requests a chunk, and 2-4) forwarding operators route the request to a storer and the chunk back to the gateway, sometimes for free or in reciprocity for services.
    5) Pairwise balances are updated, and 6) an unbalanced account is 7)rebalanced through token transfer. 
    8) Tokens are sold on the exchange layer to extract monetary rewards.
    }
    \label{fig:system_model}
\end{figure*}

Our design is based on pairwise credit accounting, with a possibility for debt forgiveness and token payments that work together in a layered architecture.  
Figure~\ref{fig:system_model} presents an overview of \tftok{} with its network, accounting and settlement layers (the exchange layer is excluded from \tftok{} but provided for completeness).

On the \textit{network layer} peers $\{p_n;n\in N\}$ form a self-organized overlay layer  using a DHT table and route requests using forwarding Kademlia, a variant of the Kademlia found in Swarm. 
Peer connections are bidirectional, meaning if peer $p_1$ is part of $p_2$'s routing table, then also $p_2$ is part of $p_1$'s routing table.

On the \textit{accounting layer}, peers maintain pairwise balances associated with each connection in the routing layer.
Pairwise balances are not verifiable to others and thus do not need transactional or enforcement mechanisms.
When receiving a chunk from a neighbor, balances are updated, and the sending neighbor is credited some accounting units.
This happens independently on every hop in the route.
We note that each peer $p_i$ performing a forwarding action is credited some amount $c_i^{in}$ from the previous peer on the route, and again credits some amount $c_i^{out}$ to the next peer. We say that the difference between these two amounts $c_i^{in} - c_i^{out}$ is the \textit{reward} $p_i$ receives for forwarding.
The reward for storing is simply the credited amount. We note that the reward for a storing action is incentivizing bandwidth usage. 
The incentivization of long-term storage is done through storage incentives, which are out of scope for this paper.

On the \textit{settlement layer}, peers transfer utility tokens to each other.
Settlements are used to rebalance pairwise accounts and may result in a monetary transfer.
We assume settlements are realized through an off-chain payment, similar to Swarm's signed receipts.
We also assume the existence of an exchange layer, where utility tokens received through a transfer can be converted to native currency.

\subsection{System Model}
We describe the system model in which the incentives operate. 
\paragraph{Normal Operation.\,}We use \tftok{} in our toolkit to study how the tokens paid by originators get distributed in a well-functioning network.
We assume no significant churn, massive failures, or massive amounts of Sybils exist in the system.

\paragraph{Selfish (rational) clients.\,}
We assume that all, or at least some gateways, are willing to make a profit in the network. Thus, peers use the maximum amount of free service and pay the requests that exceed the threshold using a token transfer.

The upload of chunks functions similarly to downloads, and can use the same layers. However, since uploads are less frequent, we focus on the retrieval of chunks. Also, in Swarm, chunk uploads are part of storage incentives.
We do not model the exchange layer or the detailed mechanisms of token transfers, 
but assume they ensure finality and incur no fee or added cost.
Such assumptions can be approximated by off-chain solutions where fees can be amortized over many individual settlements~\cite{offchain}.

We assume there is a fixed exchange rate from utility tokens to accounting units and assume that utility tokens have a stable market price reflecting their usefulness.
In the rest of this paper, we mostly ignore utility tokens. Instead, we say that a certain amount of accounting units is settled if the corresponding amount of utility tokens are being transferred.

\subsection{Debts: Forgiveness and Payments}
Reciprocity is operationalized via a pairwise credit accounting balance. 
Peers accumulate debts with their neighbors in their respective pairwise accounting balances. 
The services that peers provide to each other may balance out without requiring settlement via utility tokens. 
This way, peers can provide service without requiring settlement as long as the pairwise balance is below a given threshold.
If not, there are two possibilities: 1) once the debt hits the threshold, the indebted peer can pay tokens to reduce the debt through the settlement layer, and the creditor receives a monetary reward; 2) once the debt hits the threshold, the indebted peer can wait \lj{some time to get the debt reduced using a fixed refresh rate.}
for the refresh rate to get the debt reduced unless the creditor rejects the connection. 
If a peer always waits for 2), it receives a limited service for free.
\lj{I believe there is no point for the creditor to reject connections before refresh. If you want to add rejecting the connection, add it above where you say peers provide service.}
\lj{I added a sentence to make the connection to limited free service.}

\begin{figure}
    \centering
    \resizebox{0.5\textwidth}{!}{
    \input{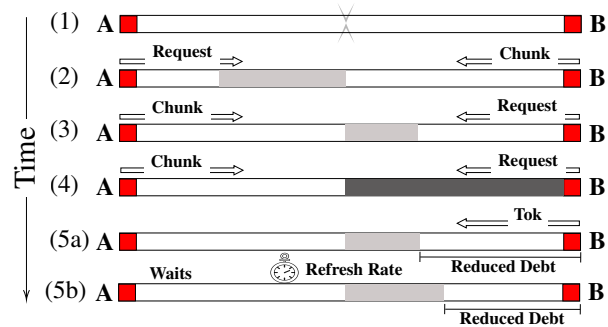}}
    \caption{Example of pairwise credit accounting balances}
    \label{fig:reciprocation}
\end{figure}

Figure~\ref{fig:reciprocation} shows an example of reciprocity. Peers A and B initialize the pairwise accounting channel with balance zero and a threshold in step $(1)$. A sends a request and goes into debt to B, and the balance tilts towards A in step $(2)$. Then B sends a request, and A provides the service, and the balance goes back around zero in step $(3)$. In step $(4)$,
B has reached the maximum debt and hits the threshold. A stops providing service until B brings the balance below the threshold again.

\section{Fairness}
\label{sec:fairness}
This section gives a primer on fairness and introduces the metrics used in this paper.

\subsection{Fairness in ODS}
The notion of fairness is generally rather broad and multifaceted, even when considered in a specific context of ODS. 
It can refer either to participants' perceived fairness of a system (and its components) or to some observable agreed-upon metric. 
While both of these aspects are important for ODS design, here we focus on fairness in the latter sense. 
\veg{I removed the last sentence - it does not tell anything.}

Wierzbicki offered an interdisciplinary view of fairness in ODS\footnote{In his book the acronym ODS refers to open distributed systems while in our paper, it refers to open decentralized systems.}.
Social psychologists judge fairness mainly through three perspectives: distributive, procedural, and retributive fairness. 
Distributive fairness is about the minimization of the differences between the shares or profits that equally entitled agents collect from a system. 
As an example, operators willing to contribute similar resources to the system, i.e., running nodes on a similar infrastructure, should receive the same token rewards.
Procedural fairness refers to impartial and fair processes, e.g., having transparent and impartial mechanisms that reward peers with honest behavior. 
Retributive fairness is about rule violations and how sanctions are in proportion with the violations, i.e., ensuring that rule-breakers are fairly treated.

Fairness also has a mathematical foundation and can be measured with some agreed metrics, as discussed later in the section.
The main focus of this paper lies in distributive fairness.
\gi{We need to explain why this type of fairness is so crucial in the context of this research, ideally already in the S.2.}

\subsection{Measuring fairness}
In the following, we define \incomefairness as a measure of distributive fairness.
\incomefairness measures how evenly the tokens, paid by gateways for downloading chunks, are distributed in the network.
For this, we determine the net income of peers by summing the payments peers receive during a given time interval and subtracting the payments performed.
In the latter, we ignore payments performed by gateways, to pay for the download of requested chunks.\footnote{Further details are provided in Appendix~\ref{app:incomefairness}.}
We then compute the Gini coefficient over the net income of different peers. While this metric of inequality has certain limitations in the context of complex macroeconomic models~\cite{yitzhaki2013gini}, it does provide helpful insights for the measurements of  decentralized systems~\cite{sai2021taxonomy}. 
The Gini coefficient is a measure of income equality with values between 0 (good) and 1 (bad).
For example, if a fraction $f$ of peers equally share all the income, the Gini coefficient is $1-f$.

Perfect \incomefairness implies that the total income received by different peers is equal. 
Assuming that every peer operator runs one peer on similar hardware, \incomefairness ensures a fair reward for making these resources available.
Assuming equal rewards per action, \incomefairness requires an equal distribution of actions to peers.
The~\incomefairness may differ with the time observed.
A few chunks requested in a large network may not be sufficient to engage all nodes and, therefore, will not generate income at all nodes. 
On the other hand, a fair distribution of income also during short periods and with small workloads indicates a more stable income and thus may better motivate nodes to stay in the system.

We note that~\incomefairness can be measured both at the accounting or settlement layer, based on whether they are applied to rewards credited on the accounting layer or settlements transferred on the settlement layer.

The Gini coefficient is a relative measure, so~\incomefairness does not provide information
about how many tokens peers actually receive.
This is well suited for our study since we assume that the value of tokens represents their utility. Thus, if more tokens are paid and received for the same service, we assume their value to be less.
\section{\tftok{} Implementation}
\label{sec:tftokimplementaion}

In this section, we give a detailed description of the accounting layer of our \tftok{} model.
We show how much credit is sent for one chunk, discuss the parameter settings for the debt threshold and limited free service layer, and discuss how and when peers transfer tokens to settle their debt.
We also discuss some challenges that arise with specific settings.

\subsection{How many accounting units for a chunk?}
\label{sec:distancerewards}
We consider two models for the credit peers receive for replying to a request: constant reward and distance-based credit.
This happens on the accounting layer of our model.
Remember that for forwarding actions, reward $c_i^{in} - c_i^{out}$ is the difference between accounting units credited to a peer $p_i$ and by $p_i$ to the next peer.
Here $c_i^{out}=c_{i+1}^{in}$

\paragraph{\textbf{Constant reward.}\,} 
In this model, all peers on the path of a request receive the same reward.
This model is not practical but is a useful baseline for our evaluation.
With constant reward, the income a peer receives is proportional to the number of answered requests.

\paragraph{\textbf{Distance-based credit.}\,} This model uses the XOR distance to find the accounting units credited for a chunk. 
The credit $c_i^{in}$ is calculated based on the distance of peer $p_i$ from the chunk.
The net credit received by $p_i$, $c_i^{in}-c_i^{out}$ is determined by the distance over which $p_i$ forwards the request and reply.
In this way, peers are motivated to forward to the peer with the shortest possible distance to the destination.
In this model, the credit sent does not depend on knowledge of the path; e.g., the credit the originator sends to the first hop in the path does not depend on the length of the path.

\begin{figure}[t]
\centering
\resizebox{0.75\textwidth}{!}{%
\begin{tikzpicture}
\tikzstyle{every node}=[font=\LARGE]
\draw [rounded corners] (5.0,13.2) rectangle (6.15,13.8);
\draw [ fill={rgb,255:red,230; green,230; blue,230} ] [rounded corners] (5.0,12.55) rectangle (7.75,13.15);
\draw [ fill={rgb,255:red,210; green,210; blue,210} ] [rounded corners]
(5.0,11.85) rectangle (9.3,12.45);
\draw [ fill={rgb,255:red,210; green,210; blue,210} ] [rounded corners]
(5.0,11.15) rectangle (9.3,11.75);
\draw [ fill={rgb,255:red,230; green,230; blue,230} ] [rounded corners] (5.0,11.17) rectangle (7.75,11.73);
\draw [ fill={rgb,255:red,255; green,255; blue,255} ] [rounded corners]
(5.0,11.19) rectangle (6.15,11.71);
\node [font=\LARGE] at (4.5,14.25) {G};
\node [font=\LARGE] at (4.5,13.55) {F1};
\node [font=\LARGE] at (4.5,12.85) {F2};
\node [font=\LARGE] at (4.5,12.15) {S};
\node [font=\LARGE] at (4.5,11.45) {C};
\node [font=\LARGE] at (5.15,14.25) {0};
\node [font=\LARGE] at (5.55,14.25) {0};
\node [font=\LARGE] at (5.95,14.25) {1};
\node [font=\LARGE] at (6.35,14.25) {1};
\node [font=\LARGE] at (6.75,14.25) {0};
\node [font=\LARGE] at (7.15,14.25) {0};
\node [font=\LARGE] at (7.55,14.25) {1};
\node [font=\LARGE] at (7.95,14.25) {1};
\node [font=\LARGE] at (8.35,14.25) {0};
\node [font=\LARGE] at (8.75,14.25) {0};
\node [font=\LARGE] at (9.15,14.25) {0};
\node [font=\LARGE] at (9.55,14.25) {0};
\node [font=\LARGE] at (9.95,14.25) {1};
\node [font=\LARGE] at (10.35,14.25) {1};
\node [font=\LARGE] at (10.75,14.25) {0};
\node [font=\LARGE] at (11.15,14.25) {0};
\node [font=\LARGE] at (11.4,14.1) {,};
\node [font=\LARGE] at (5.15,13.55) {1};
\node [font=\LARGE] at (5.55,13.55) {1};
\node [font=\LARGE] at (5.95,13.55) {1};
\node [font=\LARGE] at (6.35,13.55) {1};
\node [font=\LARGE] at (6.75,13.55) {0};
\node [font=\LARGE] at (7.15,13.55) {0};
\node [font=\LARGE] at (7.55,13.55) {1};
\node [font=\LARGE] at (7.95,13.55) {0};
\node [font=\LARGE] at (8.35,13.55) {0};
\node [font=\LARGE] at (8.75,13.55) {1};
\node [font=\LARGE] at (9.15,13.55) {1};
\node [font=\LARGE] at (9.55,13.55) {1};
\node [font=\LARGE] at (9.95,13.55) {1};
\node [font=\LARGE] at (10.35,13.55) {0};
\node [font=\LARGE] at (10.75,13.55) {0};
\node [font=\LARGE] at (11.15,13.55) {1};
\node [font=\LARGE] at (11.4,13.4) {,};
\node [font=\LARGE] at (5.15,12.85) {1};
\node [font=\LARGE] at (5.55,12.85) {1};
\node [font=\LARGE] at (5.95,12.85) {1};
\node [font=\LARGE] at (6.35,12.85) {0};
\node [font=\LARGE] at (6.75,12.85) {0};
\node [font=\LARGE] at (7.15,12.85) {1};
\node [font=\LARGE] at (7.55,12.85) {1};
\node [font=\LARGE] at (7.95,12.85) {0};
\node [font=\LARGE] at (8.35,12.85) {1};
\node [font=\LARGE] at (8.75,12.85) {0};
\node [font=\LARGE] at (9.15,12.85) {1};
\node [font=\LARGE] at (9.55,12.85) {1};
\node [font=\LARGE] at (9.95,12.85) {0};
\node [font=\LARGE] at (10.35,12.85) {0};
\node [font=\LARGE] at (10.75,12.85) {1};
\node [font=\LARGE] at (11.15,12.85) {1};
\node [font=\LARGE] at (11.4,12.7) {,};
\node [font=\LARGE] at (5.15,12.15) {1};
\node [font=\LARGE] at (5.55,12.15) {1};
\node [font=\LARGE] at (5.95,12.15) {1};
\node [font=\LARGE] at (6.35,12.15) {0};
\node [font=\LARGE] at (6.75,12.15) {0};
\node [font=\LARGE] at (7.15,12.15) {1};
\node [font=\LARGE] at (7.55,12.15) {1};
\node [font=\LARGE] at (7.95,12.15) {1};
\node [font=\LARGE] at (8.35,12.15) {0};
\node [font=\LARGE] at (8.75,12.15) {0};
\node [font=\LARGE] at (9.15,12.15) {1};
\node [font=\LARGE] at (9.55,12.15) {1};
\node [font=\LARGE] at (9.95,12.15) {1};
\node [font=\LARGE] at (10.35,12.15) {1};
\node [font=\LARGE] at (10.75,12.15) {0};
\node [font=\LARGE] at (11.15,12.15) {1};
\node [font=\LARGE] at (11.4,12) {,};
\node [font=\LARGE] at (5.15,11.45) {1};
\node [font=\LARGE] at (5.55,11.45) {1};
\node [font=\LARGE] at (5.95,11.45) {1};
\node [font=\LARGE] at (6.35,11.45) {0};
\node [font=\LARGE] at (6.75,11.45) {0};
\node [font=\LARGE] at (7.15,11.45) {1};
\node [font=\LARGE] at (7.55,11.45) {1};
\node [font=\LARGE] at (7.95,11.45) {1};
\node [font=\LARGE] at (8.35,11.45) {0};
\node [font=\LARGE] at (8.75,11.45) {0};
\node [font=\LARGE] at (9.15,11.45) {1};
\node [font=\LARGE] at (9.55,11.45) {0};
\node [font=\LARGE] at (9.95,11.45) {0};
\node [font=\LARGE] at (10.35,11.45) {1};
\node [font=\LARGE] at (10.75,11.45) {1};
\node [font=\LARGE] at (11.15,11.45) {1};
\node [font=\LARGE] at (11.4,11.3) {,};
\node [font=\LARGE] at (12.5,14.25) {$\mp=11$,};
\node [font=\LARGE] at (14,14.25) {G};
\draw [ color={rgb,255:red,224; green,27; blue,36}, line width=0.75pt, -Stealth](14.35,14.25) -- (16,14.25);
\node [font=\LARGE] at (15.15,14.55) {9};
\node [font=\LARGE] at (16.35,14.25) {F1};
\draw [ color={rgb,255:red,224; green,27; blue,36}, line width=0.75pt, -Stealth](16.65,14.25) -- (18.30,14.25);
\node [font=\LARGE] at (17.45,14.55) {5};
\node [font=\LARGE] at (18.6,14.25) {F2};
\draw [ color={rgb,255:red,224; green,27; blue,36}, line width=0.75pt, -Stealth](18.95,14.25) -- (20.60,14.25);
\node [font=\LARGE] at (19.75,14.55) {1};
\node [font=\LARGE] at (20.85,14.25) {S};
\node [font=\LARGE] at (12.5,12.85) {$\mp=16$,};
\node [font=\LARGE] at (14,12.85) {G};
\draw [ color={rgb,255:red,224; green,27; blue,36}, line width=0.75pt, -Stealth](14.35,12.85) -- (16,12.85);
\node [font=\LARGE] at (15.15,13.15) {14};
\node [font=\LARGE] at (16.35,12.85) {F1};
\draw [ color={rgb,255:red,224; green,27; blue,36}, line width=0.75pt, -Stealth](16.65,12.85) -- (18.30,12.85);
\node [font=\LARGE] at (17.45,13.15) {10};
\node [font=\LARGE] at (18.6,12.85) {F2};
\draw [ color={rgb,255:red,224; green,27; blue,36}, line width=0.75pt, -Stealth](18.95,12.85) -- (20.60,12.85);
\node [font=\LARGE] at (19.75,13.15) {6};
\node [font=\LARGE] at (20.85,12.85) {S};
\node [font=\LARGE] at (12.5,11.45) {$\mp=30$,};
\node [font=\LARGE] at (14,11.45) {G};
\draw [ color={rgb,255:red,224; green,27; blue,36}, line width=0.75pt, -Stealth](14.35,11.45) -- (16,11.45);
\node [font=\LARGE] at (15.15,11.75) {28};
\node [font=\LARGE] at (16.35,11.45) {F1};
\draw [ color={rgb,255:red,224; green,27; blue,36}, line width=0.75pt, -Stealth](16.65,11.45) -- (18.30,11.45);
\node [font=\LARGE] at (17.45,11.75) {24};
\node [font=\LARGE] at (18.6,11.45) {F2};
\draw [ color={rgb,255:red,224; green,27; blue,36}, line width=0.75pt, -Stealth](18.95,11.45) -- (20.60,11.45);
\node [font=\LARGE] at (19.75,11.75) {20};
\node [font=\LARGE] at (20.85,11.45) {S};
\end{tikzpicture}
}%
\caption{Gateway $G$ seeks chunk $C$ at depth $\delta=11$. Nodes $F1$ and $F2$ route to destination $S$. Three $\mp$ values ($\mp=11, 16, 30$) calculate $c_i^{in}$. Payment amounts on red arrows: $\mp=11$, $G$ pays $9$ to $F1$, $F1$ pays $5$ to $F2$, and retains $4$. Gray shades depict common chunk address bits.}
\label{fig:maxpo_example}
\end{figure}
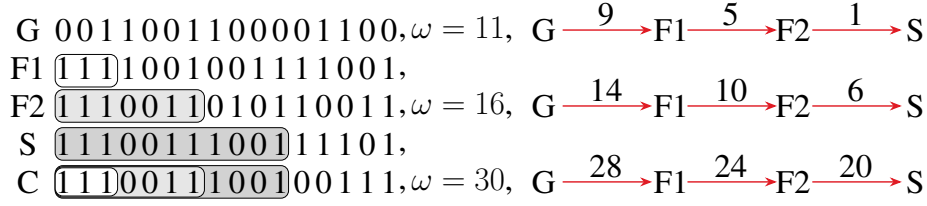

Distance-based credit is used in the Swarm network.
As the Swarm network, we use Equation~\ref{eq:maxPO} to determine $c_i^{in}$.
Here $\commonBits(p_i, \textit{chunkAddress})$ returns the number of bits in the common prefix of the two addresses. 
The constant $+1$ in the equation ensures that the last peer in the route (performing the storage action) receives at least one accounting unit.
\mp is a configurable parameter. When sending a request to peer $p_i$, $\commonBits(p_i, \textit{chunkAddress})$ is at least 1.
Thus \mp is the maximum amount for $c_i^{in}$. 
Therefore, \mp is a useful unit also for other parameters in pairwise accounting.
We note that in Swarm, $c_i^{in}$ is additionally multiplied by a constant \textit{price}, which we omit for simplicity.
\begin{equation}
\label{eq:maxPO}
    c_i^{in} = (\max(0,\mp - \commonBits(p_i, \textit{chunkAddress})) + 1)
\end{equation}
We could find no information about how to initialize $\mp$ in Swarm documents. We evaluate different parameters for $\mp$ but keep $\mp\geq\delta$, where $\delta$ is the storage depth, another parameter determining whether a peer is responsible for storing a chunk.
A peer $p_i$ is responsible to store a chunk if $\commonBits(p_i, \textit{chunkAddress})\geq \delta$.
Setting $\mp \geq \delta$ thus 
ensures $\mp > \commonBits(p_j, \textit{chunkAddress})$ for a peer $p_j$ not storing a chunk. 
Peer $p_j$ can forward the request to a different peer $p_{j+1}$ closer to the chunk, and receive $c_j^{in}-c_{j+1}^{in}>0$. 
Thus, a forwarding peer receives a non-zero net credit.
A larger $\mp$, e.g., $\mp >> \delta$ propagates a larger credit to the peers located on the last hops on the path.
Figure~\ref{fig:maxpo_example} shows an example of how many credits are sent using Equation~\ref{eq:maxPO}.

In Section~\ref{sec:evalmp} we analyze the suitable choices of the \mp parameter and show the effect distance-based rewards have on \incomefairness in Section~\ref{sec:fairaccounting}.

\subsection{Parameters for reciprocity and forgiveness}

Reciprocity is parametrized by the threshold for maximum accumulated debt.
Intuitively, a larger threshold parameter can increase reciprocity, allowing a peer first to receive multiple chunks and later repay by serving multiple requests.
However, a small threshold limits how many chunks a peer may have to provide to free-riding neighbors without receiving reciprocation.
Comparing constants in the code, we found that 
Swarm uses a constant threshold parameter on all connections, which is equal to 400 times the maximum credit sent for a single chunk, i.e., $400\times \max(c_i^{in})$.
We model this by using a constant default threshold parameter of $1 \times\max(c_i^{in})$. 
Thus, at least one chunk can be retrieved on any connection without requiring settlement.
According to Equation~\ref{eq:maxPO}, $\max(c_i^{in})=\mp$.
In this work, we are interested in how the tokens, transferred after reaching the threshold, get distributed in the network.
Thus, our model keeps the same characteristics as Swarm, but thresholds saturate faster, allowing more efficient evaluation.

Forgiveness in \tftok{} is parametrized by a \refreshrate expressed in accounting units per second. 
Thus, after $\Delta$ seconds, a peer is forgiven up to $\Delta\times \refreshrate$ many accounting units.
Note that a peer is not forgiven more than its current debt, which is limited by the threshold discussed above.
%
Swarm sets the \refreshrate allows to forgive the complete debt threshold after waiting 20 seconds.
Based on the maximum reward for a chunk, this reflects a free layer of at least 20 chunks per second on every connection.
Again, to allow more efficient evaluation, we use a smaller \refreshrate of $1/2\times \max(c_i^{in})= \mp/2$ per second. 

The introduction of reciprocity and limited free service on all connections poses a significant challenge.
We show in Section~\ref{sec:settlementfairness} that
a constant \refreshrate distributes the cost of limited free service unevenly on the path, 
impacting negatively on \incomefairness.
This is because the accounting units sent using distance-based credit, but also the frequency of use differs for different connections.
We designed a different parametrization of reciprocity and limited free service that adapts these parameters based on the 
distance between the peers adjacent to a connection.
Our evaluation shows that this pairwise parametrization can reduce the negative impact the limited free service has on \incomefairness.
Details on our pairwise adjusted threshold and \refreshrate are given in Appendix~\ref{app:adjustable_thr_forgiveness}.

\subsection{When to settle debt with tokens?}
\label{sec:settle}

When reaching the debt threshold, peers need to decide to either wait for \refreshrate and rely on the limited free service or to settle debt by transferring tokens.
We assume that all peers make use of reciprocity and the limited free service and do not settle debt unless it is required. 
Peers, therefore, try to request chunks from a neighbor, from which the pairwise balance allows retrieval without a settlement, and apply the \refreshrate whenever possible.
We also assume that peers transfer as few tokens as possible.
Thus, when requesting a chunk without free bandwidth available, a peer will only settle enough debt to be able to request the current chunk. 
We discuss other approaches to settlement in Appendix~\ref{app:paymentmodels}.

Our settlement model differs from the default settings in Swarm. According to these default settings, originators always settle their debt, while forwarding peers never do so.
If forwarding peers never transfer tokens, the limited free service imposes a limit, also on originators settling their debt, and a paying originator may have to wait for the \refreshrate on a distance hop of a requests path.
If forwarding peers settle debt with their neighbors, it allows them to serve more requests from paying neighbors and also receive more tokens.

In our simulations, both forwarding peers and originators perform settlements, but only after using reciprocity and the limited free service. 
In this setting, a paying originator is not limited by the free service layer.
However, it may happen that forwarding peers spend more tokens than they receive, resulting in \textit{negative income}.
Such negative income is due to bottlenecks in the network, where a peer may give more free bandwidth to others than it receives. 
\veg{What is a bottleneck in the network? Unclear}

In Appendix~\ref{app:paymodelimpact}, we provide measurements of the effective download rates gateways can achieve when forwarding peers do not perform settlements.
We also show that both our pairwise limited free service and especially a more balanced network can reduce the occurrence of \textit{negative income}.

It is challenging to ensure that paying originators do not have to wait for the \refreshrate while also preventing negative income, which may discourage peers from sharing their bandwidth.
We believe that this problem can be solved by more complex policies when forwarders should settle debt but leave the details for future work.


\section{Evaluation}
\label{sec:evaluation}

In this section, we evaluate the different incentive mechanisms and parameters in our \tftok{} model.
To better understand the different mechanisms, we evaluate them incrementally.
To investigate distance-based credit, we assume that no reciprocity or limited free service happens and study the distribution of accounting united.
We investigate the effect of adding first reciprocity and then limited free service.
Finally, we evaluate the effect of various additional mechanisms like caching and shuffling of connections.

%
We focus on the effect that different parameters have on 
\incomefairness, but also show other findings relevant to the configuration and effectiveness of these mechanisms.

\textbf{Parameters}
The main parameter for distance-based rewards is $\mp$, which we vary in our experiments.
As explained in Section~\ref{sec:tftokimplementaion} we bind the \textit{threshold} for maximum debt to $\mp$ and the \refreshrate to $\mp/2$.
These settings ensure that at least one chunk can be received for free over any connection every 2 seconds.
We use $\mp=16$ as default parameter, since we find that it most evenly spreads the accounting units from the originator on a single path.

Further, we explore the impact of adjusting the bucket size $k$, which determines the number of connections maintained in Kademlia. 
Our experimentation involves a network comprising $10,000$ peers. 
For comparison, in September 2023, the Swarm network contained 8457 active and 6179 staked nodes\footnote{According to \url{https://swarmscan.io/}}.
We use a storage depth $\delta=11$. This ensures that on average, four peers are responsible to store a chunk, similar as in the Swarm network.
Unless otherwise noted, peer addresses are picked uniformly at random. Thus, the actual number of peers responsible to store a chunk may vary significantly.
By default, we set the bucket size to $k=8$. This results in each peer maintaining 80 connections.
This aligns with values in the Swarm network, recently changed from 4 to 20.
Connections are used in both directions and chosen uniformly at random.

\textbf{Workload}
We use a uniform distribution of chunks, meaning that any address will be requested with the same probability.
While in real workloads, some files will be more popular than others, even popular files will contain many chunks evenly distributed among peers. 
Since our chunk addresses are mainly used to determine which peers should be contacted, we believe the assumption of uniform distribution is reasonable. 
We also did adjust a dataset showing the workload of a public IPFS gateway to chunk addresses. 
In our simulation \emph{originator} nodes represent such gateways. 
As we report in Appendix~\ref{app:ipfsdataset} (Figure~\ref{fig:ipfs_comapre_generated}), the IPFS workload gives similar results as our uniform chunks.
We change the number of peers that function as originators varying from 0.5\% (50 peers) to 100\%. 
The workload is evenly distributed among the originators. Thus, all originators request chunks at the same frequency.
We typically use a workload of 10 million chunks requested over 10 seconds.
We use larger workloads where it was necessary to make measurements converge.


\textbf{Simulation tool}
We did implement a tool\footnote{\url{https://github.com/relab/bandwidth-incentive-simulation}} that simulates routing in the Kademlia network, pairwise accounting, applying reciprocity and the limited free service, and records settlements performed.
Further details on the tool are given in Appendix~\ref{app:simulation_design}.
To mitigate the influence of randomness in our experiments, each value reported is the average over five distinct network graphs.

\subsection{Distance-based credit}

In the following, we analyze what effects distance-based credits have on balances and fairness on the accounting layer.
Distance-based credits are parametrized by the maximum price \mp.
We performed extensive simulations, varying the parameter \mp and the bucket size $k$ in peers' routing table.

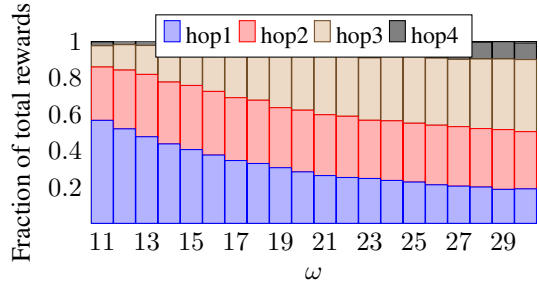
\begin{figure}[t]
    \centering
            \begin{tikzpicture}
            \pgfplotsset{width=7.5cm, height=4cm,compat=1.15}
            \def\datafileA{data/reward_over_hops_8.csv}
            \pgfplotstableread{\datafileA}\datatable
            \begin{axis}[ybar stacked,
                bar width=0.29cm,
                legend style={anchor=north,font=\small},
                xlabel={\mp},
                ylabel={Fraction of total rewards},
                xmin=11, xmax=30,
                ymin=0, ymax=1,
                xtick={11,13,15,17,19,21,23,25,27,29},
                ytick={0.2,0.4,0.6,0.8,1},
                enlarge x limits={abs=0.5*\pgfplotbarwidth},
	       	    legend entries={hop1, hop2, hop3, hop4},
	       	    legend cell align={left},
                legend style={at={(0.5, 1.15)},
	                           anchor=north,legend columns=-1},
            ]
            \addplot table [x=mp, y=hop_1, col sep=comma]{\datafileA};
            \addplot table [x=mp, y=hop_2, col sep=comma]{\datafileA};
            \addplot table [x=mp, y=hop_3, col sep=comma]{\datafileA};
            \addplot table [x=mp, y=hop_4, col sep=comma]{\datafileA};
            \end{axis}
            \end{tikzpicture}
    \caption{Fraction of rewards that goes to each hop on the route from originator to storer on average, with $k=8$, and different \mp values. Larger \mp gives a smaller fraction to the first hop.}
    \label{fig:reward_over_hops_8}
    
\end{figure}

\subsubsection{Income distribution on the path}
\label{sec:evalmp}
We investigate different parameters for $\mp$.
A larger $\mp$ means that more accounting units need to be sent, but it also changes how these units are distributed among the different peers on a path.
The distribution is more relevant to us since we assume that a higher cost in accounting units would result in a lower token value.
Figure~\ref{fig:reward_over_hops_8} shows how accounting units are distributed with varying \mp values. 

\begin{observation}
A larger \mp parameter gives a smaller fraction of the reward to the first hop in the route.
\end{observation}

\begin{figure*}[t]
    \begin{subfigure}[t]{0.32\textwidth}
        \centering
        \resizebox{\textwidth}{!}{%
            \begin{tikzpicture}
\pgfplotsset{width=7.5cm,height=6cm,compat=1.9}
\def\datafileA{data/baseincomefairnessT10sk8Th0Fg0W16.csv}
\pgfplotstableread{\datafileA}\datatable
\def\datafileB{data/baseincomefairnessT10sk8Th0Fg0W30.csv}
\pgfplotstableread{\datafileB}\datatable
\def\datafileC{data/baseincomefairnessT10sk16Th0Fg0W16.csv}
\pgfplotstableread{\datafileC}\datatable
\def\datafileD{data/baseincomefairnessT10sk16Th0Fg0W30.csv}
\pgfplotstableread{\datafileD}\datatable

\begin{axis}[
    tick label style={font=\small},
    legend style={anchor=north},
    xlabel={$\%$ of originators},
    ymin=0,
    ymax=1,
    ylabel={\incomefairness},
    xtick={1,10,50,100},
    legend pos=north east,
    ymajorgrids=true,
    legend entries={rand. k=8 $\omega=16$, rand. k=8 $\omega=30$, rand. k=16 $\omega=16$, 2choices k=8 $\omega=16$, 2choices k=8 $\omega=30$},
]
\addplot [mark=*, mark options={solid}, smooth, purple!80!white] table [x=originators, y=avg, col sep=comma]{\datafileA};
\addplot [mark=square*, mark options={solid}, smooth, blue!60!white] table [x=originators, y=avg, col sep=comma]{\datafileB};
\addplot [mark=x, black]table [x=originators, y=avg, col sep=comma]{\datafileC};
\addplot [mark=o, mark options={solid}, dashed, purple!80!white] table [x=originators, y=second, col sep=comma]{\datafileA};
\addplot [mark=square, mark options={solid}, dashed, blue!60!white] table [x=originators, y=second, col sep=comma]{\datafileB};
\end{axis}
\end{tikzpicture}
        }
        \caption{\textit{\incomefairness} with different number of originators and different \mp. Peer addresses are distributed randomly or to one of 2 choices.}
        \label{subfig:incomeFairnessXgateways}
    \end{subfigure}
    \hfill
    \begin{subfigure}[t]{0.32\textwidth}
        \centering
        \resizebox{\textwidth}{!}{%
            \begin{tikzpicture}
            \pgfplotsset{width=7.5cm,height=6cm,compat=1.9}
            \def\datafileA{data/hopincomedistributionO0T10sk_Th0Fg0W16T.csv}
            \pgfplotstableread{\datafileA}\datatable
            \def\datafileB{data/hopincomedistributionO0T10sk_Th0Fg0W30T.csv}
            \pgfplotstableread{\datafileB}\datatable
            \begin{axis}[
                tick label style={font=\small},
                legend style={anchor=north},
                xlabel={avg. hop of peer},
                ylabel={ratio of even share},
                xmin=1,
                legend pos=north east,
                ymajorgrids=true,
        		legend entries={k=8 $\mp=16$, k=8 $\mp=30$, k=16 $\mp=16$, k=16 $\mp=30$, constant k=8, constant k=16},
        		legend cell align={left}
            ]
            \addplot [mark=*, mark options={solid}, smooth, purple!80!white] table [x=hop8, y=income8n, col sep=comma]{\datafileA};
            \addplot [mark=square*, mark options={solid}, smooth, blue!60!white] table [x=hop8, y=income8n, col sep=comma]{\datafileB};
            \addplot [mark=x, black] table [x=hop16, y=income16n, col sep=comma]{\datafileA};
            \addplot [mark=star, brown] table [x=hop16, y=income16n, col sep=comma]{\datafileB};
            \addplot [mark=o, dotted, purple!80!white] table [x=hop8, y=work8n, col sep=comma]{\datafileA};
            \addplot [mark=x, dotted, black] table [x=hop16, y=work16n, col sep=comma]{\datafileA};
            \end{axis}
            \end{tikzpicture}
        }
    \caption{Correlating ratio of even income share receive with average hop of peer, for $0.5\%$ originators. Each dot represents $10\%$ of the peers. Constant variants are included.}
    \label{subfig:incomefairnessXhop}
    \end{subfigure}
    \hfill
    \begin{subfigure}[t]{0.32\textwidth}
        \centering
        \resizebox{\textwidth}{!}{%
            \begin{tikzpicture}
    \pgfplotsset{width=7.5cm,height=6cm,compat=1.9}
\def\datafileA{data/densenessincomeO100T10sk8Th0Fg0W16.csv}
\pgfplotstableread{\datafileA}\datatable

\begin{axis}[
    tick label style={font=\small},
    xlabel={$\#$peers in neighborhood},
    ylabel={ratio of even share},
    legend pos=north east,
    ymajorgrids=true,
    legend entries={k=8 $\omega=16$,k=8 $\omega=30$},
]
\addplot [mark=*, mark options={solid}, smooth, purple!80!white] table [x=denseness, y=incomew16, col sep=comma]{\datafileA};
\addplot [mark=square*, mark options={solid}, smooth, blue!60!white] table [x=denseness, y=incomew30, col sep=comma]{\datafileA};
\end{axis}
\end{tikzpicture}
        }
    \caption{Correlating ratio of even income share with the number of peers in a neighborhood, for $100\%$ originators.}
    \label{subfig:incomeXDenseness}
    \end{subfigure}
    \caption{Uneven income distribution, due to storage and forwarding discrepancy, and the impact of \mp on income faireness}
    \label{fig:incomefairness}
\end{figure*}
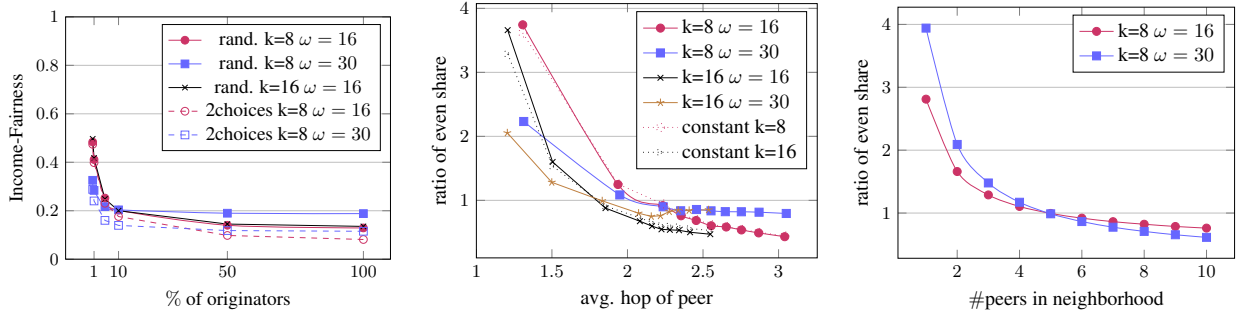

\subsubsection{\incomefairness on the accounting layer}
\label{sec:fairaccounting}
In the following, we investigate the effect of different parameters on \incomefairness.
Figure~\ref{subfig:incomeFairnessXgateways} shows the effect of the distribution of request originators (corresponding to gateways in a real life deployment) among peers.
Figure~\ref{subfig:incomeFairnessXgateways} shows \incomefairness for different values of \mp. 
It also shows the difference between networks, where peer addresses are picked uniformly at random, and a network, where every peer receives 2 choices for his address and picks the one with fewer peers within the storage distance. 
This results in a significantly more even distribution of peer addresses~\cite{twochoices}.
With $1\%$ or fewer peers as originators, income becomes significantly unequal, especially for smaller \mp (see \mp=16). With 10\%  or more peers as originators the more even peer distribution following 2 choices results in lower income fairness. 
Changing the parameter $k$ from 8 to 16 does not impact \incomefairness. 

\textbf{With 0.5\% originators} \incomefairness is 0.48 with $\mp=16$ and reduces to 0.33 with $\mp=30$. 
Figure~\ref{subfig:incomefairnessXhop} investigates the causes for this inequality.
Figure~\ref{subfig:incomefairnessXhop} shows the fraction of the total accounting units peers receive based on the average hop on which they are located on routes in the experiment. 
For example, a peer located at hop 1 for 1000 downloaded chunks and hop 2 in 2000 downloaded chunks in the experiment has an average hop of 1.66.
Accounting units are shown as the ratio of even share, where a value of 2 means that a peer receives $2\times 1/n$ of the total accounting units.
The Figure also shows values for a constant reward distribution, where every action is rewarded with a constant value. This also shows the distribution of load in the system.

As can be seen in Figure~\ref{subfig:incomefairnessXhop}, with $\mp=16$, 10\% of the peers receive $37\%$ of the total income. The constant distribution shows, that the same 10\% of peers also perform $36\%$ of the actions.
We note that a peer can only be on hop 1, if it has a connection to an originator.
With $\mp=30$ a larger share of the reward is given to the storing action, which is distributed more evenly among peers. This results in better income fairness.
While Figure~\ref{subfig:incomeFairnessXgateways} only contains data for k=8 and k=16 experiments varying k from 4 to 32 have shown no effect on the \incomefairness.

\begin{observation}
\label{obs:gateways}
Concentration of few originators results in unequal income, independent of the connectivity.
Especially, peers connected to an originator (hop 1) will receive more requests and accordingly more income.
\end{observation}

\begin{sip}
Observation~\ref{obs:gateways} suggests that Swarm should facilitate and encourage different access modes than through a web gateway (originator) to avoid an uneven load.
When uneven load cannot be avoided, a larger $\mp$ can still improve \incomefairness.
\end{sip}

\textbf{ With 100\% originators} Figure~\ref{subfig:incomeFairnessXgateways} shows a \incomefairness of 0.13 (0.19) for $\mp=16$ ($\mp=30$). This value is reduced to 0.08 (0.12) by giving peers 2 choices for their addresses. 
In Figure~\ref{subfig:incomeXDenseness} we further investigate this effect, correlating a peer's income with the number of other peers located in its neighborhood.
Peers in a densely populated neighborhood receive a smaller part of the income, since all neighbors store the same chunks.
With addresses picked uniformly at random, we see on average 58.6 out of 2048 neighborhoods with only a single peer, and 214 neighborhoods with 8 or more peers.
In the networks where peers choose their address in the sparser neighborhood out of 2 choices, we did not see neighborhoods with 1 or 8 or more peers.

\begin{observation}
    \label{obs:denseness}
    Random assignment of peer addresses leads to sparse and dense neighborhoods, which again result in unequal income.
\end{observation}

\begin{sip}
Observation~\ref{obs:denseness} suggests that peers should not simply choose their addresses at random but rather should try to achieve placement in a sparsely populated neighborhood. 
Simply using the sparser of two randomly selected neighborhoods (2choices), has a significant impact.
To the best of our knowledge, Swarm currently does not contain mechanisms to achieve such balancing.
\end{sip}
\subsection{Reciprocity and Limited free service}
\label{subsec:recipAndLimitedFreeService}
This section shows the effect reciprocity and limited free service have on income fairness at the settlement layer.
We assume that all peers perform settlements.
All measurements use distance-based credit on the accounting layer.

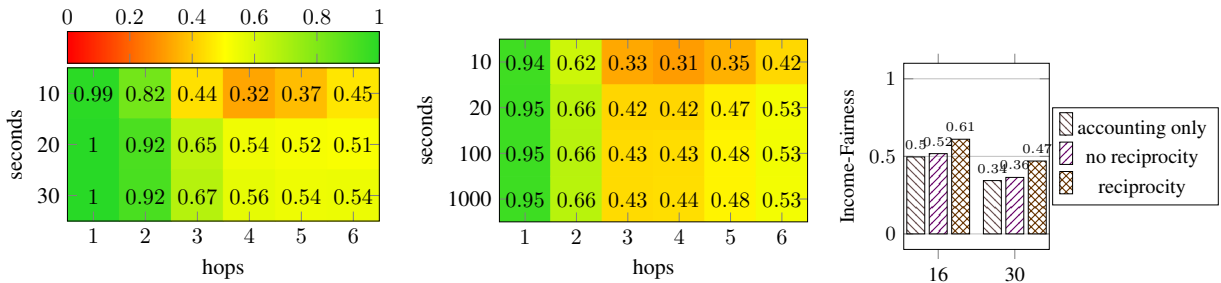
\begin{figure*}[t]
    \begin{subfigure}[b]{0.33\textwidth}
        \centering
        \resizebox{\textwidth}{!}{
        \pgfplotsset{height=4cm,width=6.5cm}
        \begin{tikzpicture}
        \begin{axis}[enlargelimits=false,
            colorbar horizontal,
            colorbar style={
                at={(0.5,1.03)},
                anchor=south,
                xticklabel pos=upper,
            },
            xlabel={hops},
            ylabel={seconds},
            point meta min=0,
            point meta max=1,
            symbolic y coords={10,20,30},
            ytick={10,20,30},
            xtick={1,2,3,4,5,6},
            colormap={reverse}{color(0)=(red) color(0.5)=(yellow) color(1)=(green!70!gray)}]
            \addplot [
                matrix plot,
                nodes near coords,
                node near coords style={yshift=-0.25cm},
                point meta=explicit,
            ] table [meta expr=\thisrow{C}, col sep=comma] {data/hoppayedpercentnorecO05.csv};
        \end{axis}
        \end{tikzpicture}    
        }
        \caption{Fraction of paid forwarded chunks with no reciprocity and $\mp=16$, using a fixed debt threshold and direction.}
        \label{subfig:payfractionhonNorec}
    \end{subfigure}
    \begin{subfigure}[b]{0.33\textwidth}
        \centering
        \resizebox{\textwidth}{!}{%
        \pgfplotsset{height=4.5cm,width=6.5cm}
        \begin{tikzpicture}
            \begin{axis}[enlargelimits=false,
                ylabel={seconds},
                xlabel={hops},
                point meta min=0,
                point meta max=1,
                symbolic y coords={10,20,100,1000},
                ytick={10,20,100,1000},
                xtick={1,2,3,4,5,6},
                colormap={reverse}{color(0)=(red) color(0.5)=(yellow) color(1)=(green!70!gray)}]
                \addplot [
                    matrix plot,
                    nodes near coords,
                    node near coords style={yshift=-0.25cm},
                    point meta=explicit,
                ] table [meta expr=\thisrow{C}, col sep=comma] {data/hoppayedpercentreciprocityO05.csv};
            \end{axis}
            \end{tikzpicture}    
        }
        \caption{Fraction of paid forwarded chunks with $\mp=16$ with reciprocity, allowing to exchange bandwidth for bandwidth.}
    \label{subfig:payfractionhonRec}
    \end{subfigure}
    \hfill
    \begin{subfigure}[b]{0.32\textwidth}
        \centering
        \resizebox{\textwidth}{!}{%
            \begin{tikzpicture}
    \pgfplotsset{width=4cm, height=4.5cm,compat=1.9}
    \begin{axis}[
        tick label style={font=\small},
        label style={font=\small},
        legend style={anchor=north,font=\small},
        symbolic x coords={16, 30},
        xtick=data,
        x=1.2cm,
        nodes near coords,
        nodes near coords style={font=\tiny},
	    enlarge y limits=0.1,
        enlarge x limits=0.45,
	    ybar,
        ymin=0, ymax=1,
        legend image code/.code={
            \draw [#1] (0cm,-0.1cm) rectangle (0.2cm,0.25cm); },
        xlabel={\mp},
        ylabel={\incomefairness},
        legend style={at={(1.02, 0.5)},
	    anchor=west,legend columns=1},
        bar width = 8pt,
        ymajorgrids=true,
        cycle list={
                    {pattern=north west lines, pattern color=pink!40!black},
                    {pattern=north east lines, pattern color=magenta!40!black},
                    {pattern=crosshatch, pattern color=orange!40!black},
                },
    ]
    \addplot coordinates {(16, 0.4962858) (30,0.3437794)};
    \addplot coordinates {(16, 0.517485) (30,0.364232)};
    \addplot coordinates {(16, 0.6103876) (30,0.4697268)};
    \legend{accounting only, no reciprocity, reciprocity}
    \end{axis}
\end{tikzpicture}
        }
        \caption{\incomefairness with $0.5\%$ gateways, with and without reciprocity.}
        \label{subfig:incomeFairnessReciprocityO05}
    \end{subfigure}
    \caption{\textit{Reciprocity and \incomefairness} The effect of reciprocity on \incomefairness with $0.5\%$ originators, each requesting $2,000$ chunks per second.
    Heatmaps show the saturation of edges with and without reciprocity.}
    \label{fig:reciprocity}
\end{figure*}

The limited free service allows peers to download some chunks for free every timestep.
When measuring fairness on the settlement layer, it is therefore important to vary not only the number of chunk download requests raised to the system but also the rate at which these requests are issued.
We let every originator to retrieve between $10$ and $2,000$ chunks per timestep.
Assuming a chunk size of 4kB as in Swarm these numbers result in a request rate between $40kB/s$ and $8MB/s$.

\subsubsection{Distributing the cost of free service on the path}
\label{sec:settlementfairness}
In the following, we investigate the effect reciprocity and the free service layer have on \incomefairness.

In Figure~\ref{fig:reciprocity} we investigate reciprocity without applying the \refreshrate from the limited free service. 
Without \refreshrate, peers still go into debt with each other. 
The threshold for maximum debt is set to $\mp$. 
With reciprocity, peers balance out debt given to each other.

To better understand the effect of reciprocity, we also implemented a variant, where peers can go into debt with each other, until the threshold is reached, but the pairwise debts of neighbors are not substracted from each other. 
We refer to this variant as \textit{no reciprocity}.
With no reciprocity, connections hit the threshold significantly faster than with reciprocity, as can be seen from Figure~\ref{subfig:payfractionhonNorec} and~\ref{subfig:payfractionhonRec}.

\begin{observation}
    \label{obs:reciprocityunfair}
    Reciprocity worsens \incomefairness since requests on the first hop more often produce settlements.
\end{observation}

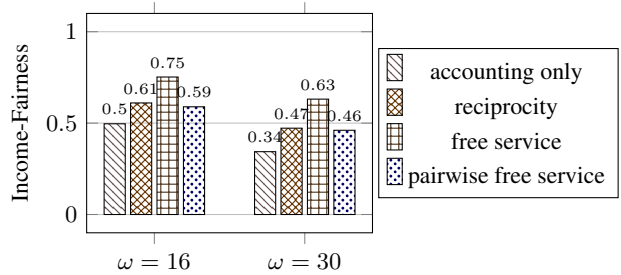
\begin{figure}[t]
    \centering
    \begin{tikzpicture}
    \pgfplotsset{width=6cm, height=4.5cm,compat=1.9}
    \begin{axis}[
        tick label style={font=\small},
        label style={font=\small},
        legend style={anchor=north,font=\small},
        symbolic x coords={16, 30},
        xtick=data,
        xticklabels={$\mp=16$, $\mp=30$},
        x=2cm,
        nodes near coords,
        nodes near coords style={font=\tiny},
	    enlarge y limits=0.1,
        enlarge x limits=0.45,
	    ybar,
        ymin=0, ymax=1,
        legend image code/.code={
            \draw [#1] (0cm,-0.1cm) rectangle (0.2cm,0.25cm); },
        ylabel={\incomefairness},
        legend style={at={(1.02, 0.5)},
	    anchor=west,legend columns=1},
        bar width = 8pt,
        ymajorgrids=true,
        cycle list={
                    {pattern=north west lines, pattern color=pink!40!black},
                    {pattern=crosshatch, pattern color=orange!40!black},
                    {pattern=grid, pattern color=brown!40!black},
                    {pattern=crosshatch dots, pattern color=blue!40!black},
                },
    ]
    \addplot coordinates {(16, 0.4962858) (30,0.3437794)};
    \addplot coordinates {(16, 0.6103876) (30,0.4722798)};
    \addplot coordinates {(16, 0.7522242) (30,0.6312535)};
    \addplot coordinates {(16, 0.589346) (30,0.4608434)};
    \legend{accounting only, reciprocity, free service, pairwise free service}
    \end{axis}
\end{tikzpicture}
    \caption{\textit{Limited free service and \incomefairness} 
        The effect of limited free service on \incomefairness with $0.5\%$ originators. 
        Variants included: accounting only considers only amounts credited on the accouting layer; reciprocity does not provide free service, setting the \refreshrate to zero; free service uses the \refreshrate $\mp/2$; pairwise free service, adapts threshold and \refreshrate to peer distance.}
    \label{fig:income-fairness-refresh-05gateway}
\end{figure}

In Figure~\ref{fig:income-fairness-refresh-05gateway}, we investigate the effect of the limited free service on \incomefairness.
We also show results of our \textit{pairwise limited free service}, where threshold and \refreshrate are set based on the proximity of adjacent peers.
We ensure the adjusted threshold is still larger than the accounting units required for any chunk forwarded on that connection.
More details on how we adjust \textit{threshold} and \refreshrate can be found in Appendix~\ref{app:adjustable_thr_forgiveness}.

Figure~\ref{fig:income-fairness-refresh-05gateway} shows that limited free service can further worsen \incomefairness, compared to using only reciprocity.
While Figure~\ref{fig:income-fairness-refresh-05gateway} shows results for $0.5\%$ gateways, a larger number of gateways results in similar results (see Appendix~\ref{app:100gateways}).

\begin{observation}
    The default variant of limited free service significantly reduced \incomefairness.
\end{observation}

Our pairwise limited free service mitigates this difference, resulting in similar \incomefairness as the variant not providing free service. 
This shows that our pairwise limits distribute the cost of free service more equally among peers on a path.

\begin{observation}
\label{obs:adjustable_model}
    Our pairwise limited free service allows significantly improves \incomefairness.
\end{observation}

\begin{sip}
    Following Observation~\ref{obs:adjustable_model}, Swarm could achieve a better \incomefairness by introducing our adaptive free service limit.
\end{sip}

\subsubsection{Centralization risk through gateways}
\label{sec:clique}
According to Observation~\ref{obs:gateways}, peers adjacent to an originator, generating a lot of requests as a gateway, receive more income.
Accordingly, such originators can reduce their 
overall cost by connecting to other originators who generate a lot of requests.
Table~\ref{tab:cluster} shows how a cluster of 5 or 100 originators can reduce its cost, and how many of the hops are performed between originators.
With an average path length of 2.6 hops, if 45\% of hops are performed between 100 originators, other peers are mostly used for the retrieval of chunks but not for their forwarding.
Enabling reciprocity further increases this effect, since peers prefer to exchange bandwidth for bandwidth, rather than settling with tokens.

\begin{table*}
    \centering
    \caption{Centralization risk through originator cliques}
    \label{tab:cluster}
    \begin{tabular}{|l|c|c|c|c|}
        \hline
        \multicolumn{1}{|r|}{\#\textit{originators}} & 5 & 100 & 5 & 100 \\
        & \multicolumn{2}{|l|}{tok per chunk} &\multicolumn{2}{|l|}{internal hops} \\
        \hline
        random with recip. & 11.6 & 11.1 & 0.1\% & 1.6\% \\        
        clique no recip. & 11.0 & 7.0 & 5.9\% & 38.4\% \\        
        clique with recip. & 10.1 & 6.5 & 17.5\% & 45.8\% \\
        external clique & 9.7 & 5.9 & -- & -- \\
        \hline
    \end{tabular}
\end{table*}

This shows that originators are incentivized to form clusters, monopolizing a large fraction of the traffic. 
We also show the cost an operator of multiple gateways, acting as originators, can achieve by requesting each chunk from the originator closest to the chunk address. 
In this variant, the originators 
collude through external connection without pairwise accounting to form an (\textit{external clique}).
Table~\ref{tab:cluster} shows that by connecting inside the system, gateway operators can achieve similar benefits as an operator of multiple originators.


\begin{observation}
Originators are incentivized to form clusters but do not gain significant benefits from external cliques.
\end{observation}

\subsubsection{Caching}
\label{sec:caching}
Forwarding Kademlia allows peers to cache chunks they forward to an originator.
We did implement such caching and evaluate its effect on \incomefairness. 
Since caching requires a non-uniform workload, we use the adapted IPFS workload for this evaluation.

Our experiments detailed in Appendix~\ref{app:caching} show that with few originators caching may increase \incomefairness from 0.57 to 0.71. However, if the cache at originators is very large, \incomefairness reduces again (0.67), since mostly unique chunks are requested from the network.

With many originators (e.g., 20\%) caching may have positive effect, e.g. improving fairness from 0.26 to 0.23. Caching can reduce the imbalance caused by neighborhoods with few peers, reducing the ratio of tokens peers in a neighborhood of size 1 receive from 3.3 to 2.9 times the fair share.

\begin{observation}
Caching can smoothen imbalances related to the peer distribution in the address space, but increases imbalances with few originators.
\end{observation}
\subsection{Income Distribution Across Path}

One of the reasons behind unfairness is the static routing table in the underlying DHT. The limited number of neighbors, together with a condensed source of requests puts the nodes on the top buckets of the originators in an advantageous position. The topmost bucket of originators is responsible for handling half of the requests while containing only $k$ nodes.

\begin{figure}[t]
    \centering
    \resizebox{0.5\textwidth}{!}{%
        \begin{tikzpicture}
\pgfplotsset{height=6cm, width=8cm,compat=1.9}
\def\datafileA{data/shuffling.csv}
\pgfplotstableread{\datafileA}\datatable

\begin{axis}[
    tick label style={font=\footnotesize},
    legend columns = 1,
    label style={font=\footnotesize},
    legend style={anchor=north,font=\scriptsize},
    xlabel={seconds},
    ymin=0,
    ymax=1,
    ylabel={\incomefairness},
    xtick={1, 25, 50, 75, 100},
    legend pos=north east,
    ymajorgrids=true,
    legend entries={No shuffling,
    Shuffling originator neighbors,
    Shuffling neighbors,
    Originator's $k = 8$,
    Originator's $k = 256$
    }
]

\addlegendimage{mark=*, color=purple!80!gray}
\addlegendimage{mark=*, color=blue!60!black}
\addlegendimage{mark=*, color=magenta!80!gray}
\addlegendimage{style=solid}
\addlegendimage{style=dashed}

\addplot [mark=*, mark options={solid}, smooth, purple!80!gray] table [x=Time, y=NoneFairness, col sep=comma]{\datafileA};
\addplot [mark=*, mark options={solid}, smooth, blue!60!black] table [x=Time, y=GatewayEdgeFairness, col sep=comma]{\datafileA};
\addplot [mark=*, mark options={solid}, smooth, magenta!80!gray] table [x=Time, y=EdgeFairness, col sep=comma]{\datafileA};

\addplot [mark=o, mark options={solid}, dashed, purple!80!gray] table [x=Time, y=NoneFairness256, col sep=comma]{\datafileA};
\addplot [mark=o, mark options={solid}, dashed, blue!60!black] table [x=Time, y=GatewayEdgeFairness256, col sep=comma]{\datafileA};
\addplot [mark=o, mark options={solid}, dashed, magenta!80!gray] table [x=Time, y=EdgeFairness256, col sep=comma]{\datafileA};
\end{axis}
\end{tikzpicture}
    }

\caption{\incomefairness of different shuffling variants for $0.5\%$ originators, each requesting $2,000$ chunks per second with $\mp=16$, pairwise limited free service and reciprocity enabled.}
\label{fig:shuffling}
\end{figure}
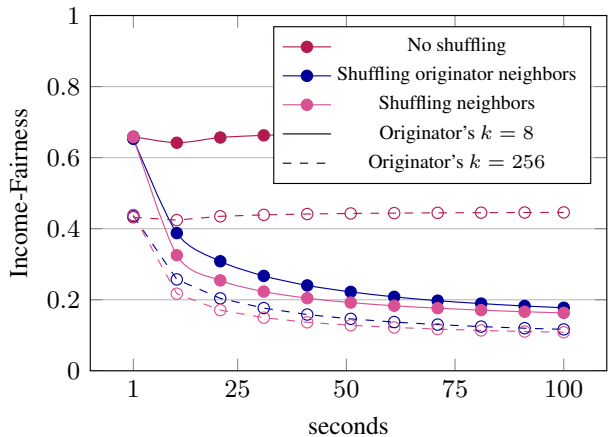

Figure~\ref{fig:shuffling} shows that if originators connect to  significantly more nodes ($k=256$), income fairness decreases by $0.2$. However, this solution may not scale since it requires originators to maintain thousands of connections.
Instead, we propose a shuffling mechanism where peers regularly change their neighbors. Thus, peers can have a large number of neighbors distributed over time.
Appendix \ref{app:shuffling} discusses how shuffling can be implemented and the side effects this has.

To evaluate how shuffling works, we considered two variants of shuffling: 1. Only originators shuffle their neighbors, 2. All nodes shuffle their neighbors. We ran these variants in the \tftok{} model with reciprocity and pairwise limited free service.
Figure~\ref{fig:shuffling} shows how shuffling affects the \incomefairness in the network.

We observe that shuffling the neighbors of originators eventually converges to shuffling every node's neighbors. During our experiment, we found that shuffling originator's neighbors results in a $13.6\%$ reduction in the cost of each chunk since originators may abandon connections with debt.

\section{Conclusion}
\label{sec:conclusion}

In this work, we present \emph{Tit-for-Token}, a framework to study token-based incentives across accounting and settlement layers of decentralized storage systems. 
We propose the triad of altruism, reciprocity, and monetary incentives as a compound of worthwhile incentive mechanisms and study the interplay between these incentives and the storage-, and network-parameters. 
We quantify income-fairness using multiple model instantiations and propose effective methods to reduce inequalities introduced by gateways. 
The Tit-for-Token framework can help system designers to improve the design of incentive mechanisms.

\section*{Acknowledgements}
We express our gratitude to the following individuals and organizations for their valuable contributions and support during the course of this research:
\begin{itemize}
    \item Filip B. Gotten, Rasmus Øglænd, and Torjus J. Knudsen, whose dedicated work significantly improved our simulation tool as part of their bachelor theses.
    \item The Swarm community, with special appreciation to Daniel Nagy, for their insightful discussions on an early version of this work.
    \item Derouich Abdessalam for his valuable contribution and feedback concerning the shuffling solution.
\end{itemize}

We also acknowledge the financial support provided by the BBChain project, granted under the Research Council of Norway (grant 274451).
\newpage
\bibliographystyle{plain}
\bibliography{references.bib}
\begin{appendix}

\section{Routing table in Kademlia}
\begin{figure}
        \centering
        \includegraphics[width=0.75\textwidth]{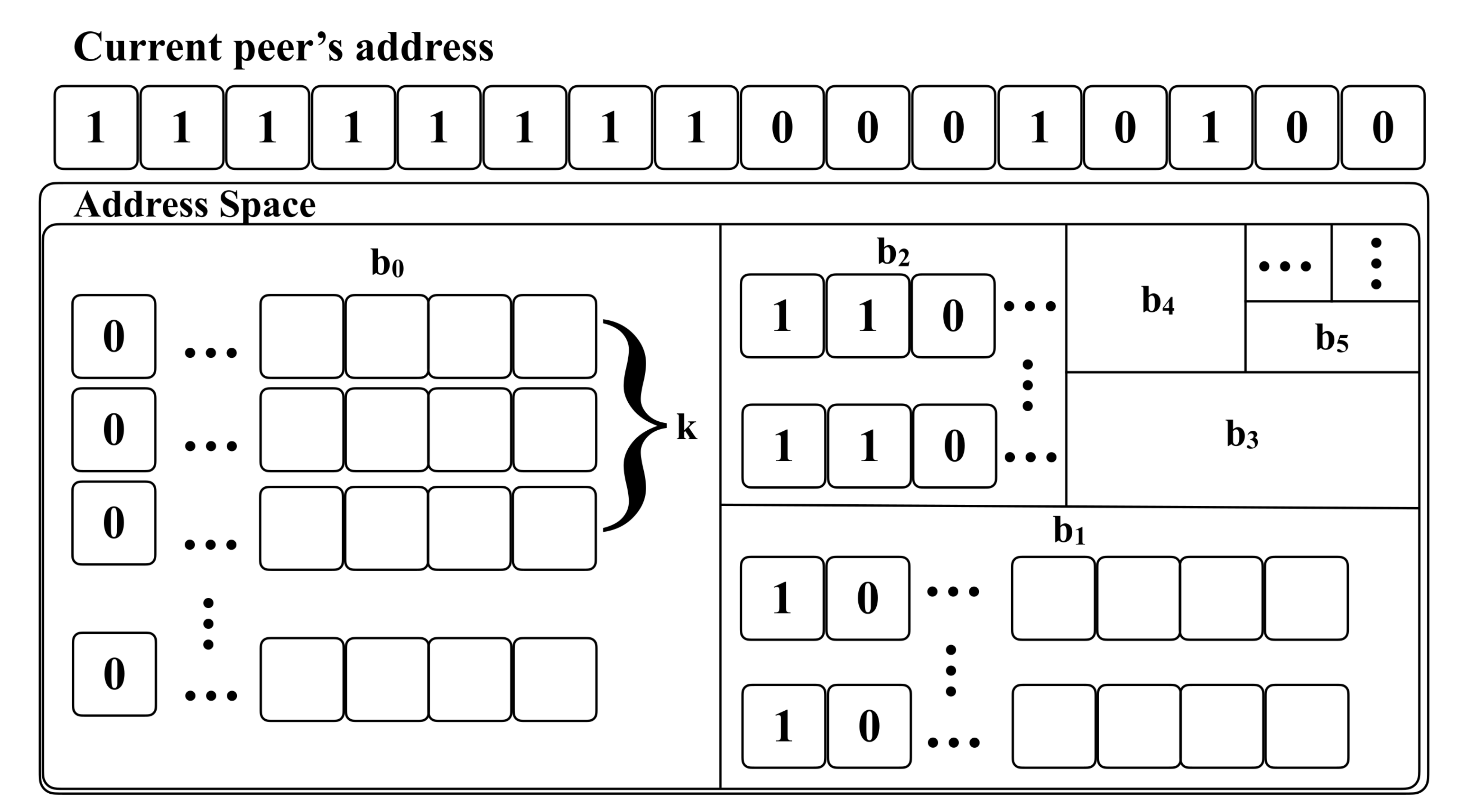}
        \caption{An example for the routing table and buckets.}
        \label{fig:appendix_routin_table}
    \end{figure}

    Figure~\ref{fig:appendix_routin_table} illustrates the process of creating a routing table. In the top row, the address of the current node is shown. Assuming that the entire address space is a rectangle and the addresses are uniformly spread in ascending order, the rectangle is divided into two halves by changing the first bit. The first half contains all nodes with the first bit zero, while the second half includes nodes with the first bit one. Since the first bit of the current peer is one, all candidates for bucket zero are found in the other half. The same procedure is repeated for the second bit, with peers having the same first bit but different second bits considered as candidates for bucket one. This process continues until the last bit, which further divides the remaining rectangles into two parts. With the bucket size denoted as k, peers will be added to each bucket by a peer. For simplicity, it is assumed that a 
    peer can see the whole network and create the routing table, however, in reality, peers have a limited view of the network and create the 
    routing table by contacting other peers.
    \label{app:A_routing_table}


\section{Computing \incomefairness}
    \label{app:incomefairness}
    Here we give a precise formula for the calculation of the \incomefairness.
    We assume that each successful request results in the retrieval of one chunk, an update of p2p balances and possibly settlement payments. 
    Assume during a time interval $t$, chunks $W=\{w_1,w_2,\ldots\}$ are successfully routed through the system.
    Let $x^{in}_{n,w}$ denote the settlement that $p_n$ receives during the routing of $w\in W$.
    Let $x^{out}_{n,w}=0$ if $p_n$ is the first peer (gateway) on the route of $w$.
    Let $x^{out}_{n,w}$ denote the settlement that $p_n$ performs during the routing if it is not the first node in the route.

    \begin{description}
        \item[\incomefairness] is the Gini-coefficient over the total income of different peers during the routing of chunks $W$:
        $$\textit{\incomefairness}(W)=\gini(\{|\sum_{w\in W}x^{in}_{n,w}-x^{out}_{n,w}|; n \in N\})$$
    \end{description}
    The Gini-Coefficient can be computed as:
    \begin{equation*}
    \gini(\{m_i; i \in N\})=\frac{\sum_{i,j\in N} |m_i-m_j|}{ 2 |N|\cdot \sum_{i\in N}m_i}
    \end{equation*}

\section{Incentives Toolkit}
\label{app:simulation_design}

\begin{figure*}
\centering
    \includegraphics[width=0.8\textwidth]{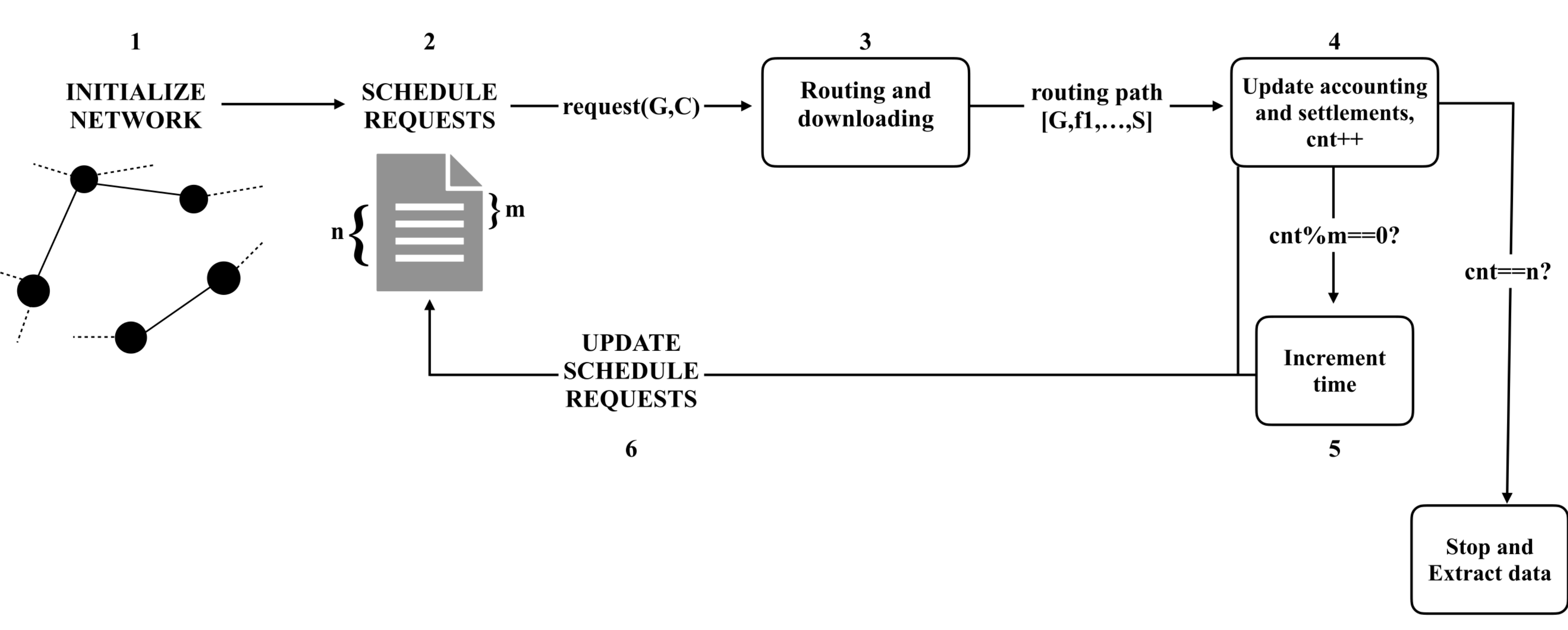}
    \caption{Overview of the simulation tool's structure in $6$ steps. Outgoing arrows with a question mark show the conditions applied to that output. Other outgoing arrows are data edges, including the output data as an input for the next component.}
    \label{fig:simulation_structure}
\end{figure*}

We implemented a modular toolkit that allows to evaluate a large range of incentive settings.
We have successfully simulated networks with up to $100,000$ peers.
The tool is written in Go, employs multithreading and simulates the download of $10$ million chunks in less than a minute.
Figure~\ref{fig:simulation_structure} demonstrates a high-level overview of the design and components of our tool.
Steps $1$ to $6$ are explained in detail in the following.

We designed and implemented our simulation tool with three general layers: the initialization layer, the policy functions layer, and the update functions layer. These three layers are decoupled so that changing a layer does not affect other layers and is easy to modify. The first step in Figure~\ref{fig:simulation_structure} is INITIALIZE NETWORK. In this step, with the input parameters generated by the user in the form of a file, a p2p network is created. These parameters include network-specific parameters such as network size, number of gateways, bits of address space, $k$, etc., and simulation parameters such as the number of chunk downloads in total, number of requests per second, etc. The design of the parameter adoption is flexible, e.g., assigning different $k$ values to different buckets is possible. Peers create their routing table and are ready to work. It continues with SCHEDULING REQUESTS in step $2$. A chunk download request is a pair of two addresses containing the gateway's address \textbf{G} and the desired chunk's address \textbf{C}. We assume chunks are uniformly distributed in the network with their addresses. The request scheduler generates $n$ pairs of requests ready to be executed. Then download requests go one by one for execution to step $3$. For each epoch of simulation $cnt$, one request goes to the Routing and downloading step. Using the forwarding Kademlia, the gateway downloads the chunk, and the output of this step is a routing path showing the participants in the downloading process. The generated routing path containing the address of the gateway, forwarders, and storer goes to step $4$. Update functions take the routing path generated in the previous step, update p2p accountings and settlements using it, and increase the $cnt$. In this step, the simulation checks for the end condition, and it $cnt$ equals the number of all requests $n$, ends the simulation and extracts the data. In order to have a notion of time and concurrency in the simulation, we define variable $m$ at the beginning. We use the $m$ variable to increase the time. With $m$, for each time unit in the simulation, $m$ requests are executed. In step $6$, the request scheduler updates the list of requests and continues with another chunk download request.


    \section{A pairwise limit for free service}
    \label{app:adjustable_thr_forgiveness}
    Through our experiments, we have observed that connections located in the farthest buckets in the Kademlia routing table seldom become saturated. For example, in Figure~\ref{subfig:payfractionhonRec}, we see that the connections in hop one get saturated more quickly. 
    This results in an uneven distribution of the cost of providing limited free service. 
    Peers on the hops further away from the originator end up providing free chunks for which the peer at hop 1 receives tokens.
    On the other hand, an originator relying only on the limited free service can download many chunks from its immediate vicinity, but only a few located far away in the address space. Thus, the gateway can quickly download half a file, but will wait long time for the last chunk.
    We identified that the cause for this is that we use the same parameters for the debt threshold and \refreshrate on all pairwise balances, resulting in a constant limit for the free service.
    While the Swarm network aims to allow peers to adapt these parameters on a pairwise basis, their default setting is still relevant.
    In the following we show how the debt threshold and \refreshrate can be initialized based on peers distance and argue why this results in a fairer distribution of costs.

    \begin{table}[t]
        \caption{Accouting units credited per chunk in different buckets and example of threshold and forgiveness based on peers' distance}
        \label{tab:bucketaccouting}
        \centering
        \begin{tabular}{|c|c|c|c|}
            \hline
            bucket & avg. credit & \multicolumn{2}{|l|}{adaptive} \\
            & & threshold & refresh rate \\
            \hline
            0 & 11.558 & 16 & 8 \\
            1 & 10.556 & 15 & 3 \\
            2 & 9.604  & 14 & 3 \\
            3 & 8.564  & 13 & 3 \\
            4 & 7.581  & 12 & 3 \\
            5 & 6.588  & 11 & 2 \\
            6 & 5.651  & 10 & 2 \\
            7 & 4.700  & 9  & 2 \\
            8 & 3.916  & 8  & 2 \\
            9 & 3.611  & 7  & 1 \\
            \hline
        \end{tabular}
    \end{table}
    
    \subsection{Adapting debt threshold based on peers' distance}
    One reason for unsaturated connections in higher buckets is that the credit sent for one chunk, depends largely on the distance between peers and accordingly, the bucket in which the peers' connection lies.
    Table~\ref{tab:bucketaccouting} shows how the average credit sent for a chunk decrease with increasing buckets. 
    Our default threshold of 16 covers less than two requests on buckets 0, while it covers 4 requests in bucket 8.
    
    This is a result of distance-based credit and we derive the theoretical maximum price for a chunk transfered between two peers below.
    If a peer $p_i$ forwards a request to peer $p_{i+1}$, then $p_{i+1}$ is located closer to the requested chunk. Especially the following equations hold:
    \begin{equation}
        \commonBits(p_{i+1},chunkId) > \commonBits(p_i,chunkId) = commonBits(p_i,p_{i+1})
    \end{equation}

    \noindent
    Thus, according to Equation~\ref{eq:maxPO}, the accounting units credited per request on 
    the connection between $p_i$ and $p_{i+1}$ is bounded by: 
    \begin{equation}
        c_{i+1}^{in} = (\mp-\commonBits(p_{i+1},chunkId) +1) \leq (\mp-\commonBits(p_i,p_{i+1}) +1)
    \end{equation}
    We use this bound as \textit{adapted threshold}.

    \begin{table}[t]
        \caption{Frequency of connection use in certain buckets on gateway and complete graph.}
        \label{tab:bucketuse}
        \centering
        \begin{tabular}{|c|c|c|}
            \hline
            bucket & use on gateway & overall use \\
            \hline
            0 & 1.000 & 1.000 \\
            1 & 0.500 & 0.509 \\
            2 & 0.250 & 0.345 \\
            3 & 0.125 & 0.451 \\
            4 & 0.062 & 0.454 \\
            5 & 0.031 & 0.468 \\
            6 & 0.016 & 0.455 \\
            7 & 0.008 & 0.448 \\
            8 & 0.004 & 0.456 \\
            9 & 0.002 & 0.454 \\
            10& 0.001 & 0.462 \\
            \hline
        \end{tabular}
    \end{table}

    \paragraph{Adapting \refreshrate based on peers' distance}
    We can adapt the \refreshrate based on our threshold, following Equation~\ref{eq:adaptthresh}
    
    \begin{equation}
        \label{eq:adaptthresh}
        \textit{adapted refresh rate} = \refreshrate \cdot \frac{\textit{adapted threshold}}{\textit{threshold}}
    \end{equation}
    However, we found this is insufficient to balance the available free bandwidth.
    We find that the second reason that connections in higher buckets do not become saturated is that these connections are used less frequently.
    Table~\ref{tab:bucketuse} shows how frequently connections in different buckets are used. Numbers are normalized with respect to the use of bucket 0.
    We see that bucket 0 is used approximately twice as often as other buckets.
    We therefore additionally half the new \refreshrate in all other buckets.

    We confirmed that for peers not performing settlements, our pairwise adapted \refreshrate gives similar limited free service as the constant threshold. This is because we keep the same threshold and \refreshrate on bucket 0, where peers send most of their requests.

\section{Understanding distance based credits}
In this section we report on additions experiments that help to understand distance based credits and the effects of the parameter $\mp$.

\subsubsection{Rewarding short routes}
Using distance-based credits, forwarding peers receive net credit proportional to how far they have forwarded a request.
Table~\ref{tab:mean_k_related_forward} shows the average accounting units a peer receives for performing a forwarding action using distance-based credit. 
The measurement uses our default parameter $\mp=16$.
Numbers show forwarders receive a larger reward in networks with a larger bucket size of $k$.
With larger $k$, peers maintain a larger routing table giving a larger choice of the next hop.
Table~\ref{tab:mean_k_related_forward} also shows the average path length from the same experiment.
A comparison of the two columns shows that the longer forwarding distances result in shorter paths.
\begin{observation}
Forwarders earn more rewards for finding shorter routes.
\end{observation}

\begin{table}[t]
    \centering
    \caption{\textit{Path length}: Routing over a larger distance gives shorter routes and larger rewards for forwarders.}
    \begin{tabular}{|c|c|c|}
    \hline
    k & Mean reward per forward & Average \# of hops \\
    \hline
    4 &  3.293  &  2.8548 \\ 
    8 &  4.101  &  2.6324 \\
    16 & 4.673  &  2.3978 \\
    20 & 4.931  &  2.3135 \\
    \hline
    \end{tabular}
    \label{tab:mean_k_related_forward}
\end{table}

  \section{\incomefairness and free service with 20\% originators}
  \label{app:100gateways}
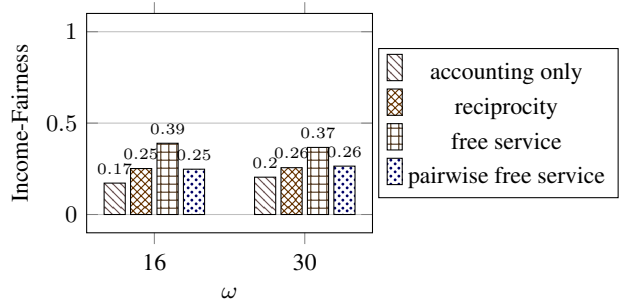
\begin{figure}[t]
    \centering
    \begin{tikzpicture}
    \pgfplotsset{width=6cm, height=4.5cm,compat=1.9}
    \begin{axis}[
        tick label style={font=\small},
        label style={font=\small},
        legend style={anchor=north,font=\small},
        symbolic x coords={16, 30},
        xtick=data,
        x=2cm,
        nodes near coords,
        nodes near coords style={font=\tiny},
	    enlarge y limits=0.1,
        enlarge x limits=0.45,
	    ybar,
        ymin=0, ymax=1,
        legend image code/.code={
            \draw [#1] (0cm,-0.1cm) rectangle (0.2cm,0.25cm); },
        xlabel={\mp},
        ylabel={\incomefairness},
        legend style={at={(1.02, 0.5)},
	    anchor=west,legend columns=1},
        bar width = 8pt,
        ymajorgrids=true,
        cycle list={
                    {pattern=north west lines, pattern color=pink!40!black},
                    {pattern=crosshatch, pattern color=orange!40!black},
                    {pattern=grid, pattern color=brown!40!black},
                    {pattern=crosshatch dots, pattern color=blue!40!black},
                },
    ]
    \addplot coordinates {(16, 0.1718124) (30, 0.2042794)};
    \addplot coordinates {(16, 0.2515106) (30, 0.2561844)};
    \addplot coordinates {(16, 0.3898528) (30, 0.3670742)};
    \addplot coordinates {(16, 0.2479358) (30, 0.264639)};
    \legend{accounting only, reciprocity, free service, pairwise free service}
    \end{axis}
\end{tikzpicture}
    \caption{\textit{Limited free service and \incomefairness} 
        The effect of limited free service on \incomefairness with $20\%$ gateways, each requesting $50$ chunks per second. 
        \incomefairness is measured only among peers not originating requests.        
Variants included: accounting only considers only amounts credited on the accounting layer; reciprocity does not provide free service, setting the \refreshrate to zero; free service uses the \refreshrate $\mp/2$; pairwise free service, adapts threshold and \refreshrate to peer distance.
}
    \label{fig:income-fairness-refresh-100gateway}
\end{figure}

In this section, we show the effectiveness of the adaptive threshold with a larger number of peers functioning as originators. In Figure~\ref{fig:income-fairness-refresh-100gateway} $20\%$ of the peers function as originators and request chunks.
We note that with reciprocity and limited free service, it is no longer meaningful to separate the payments performed by an originator into payments done while forwarding others' requests, and payments done for downloading chunks. Some originators may receive less income from forwarding, but also perform fewer settlements per request.
Therefore, Figure~\ref{fig:income-fairness-refresh-100gateway} shows \incomefairness only among the forwarding peers, not requesting chunks.
We see that allowing an equal free service on all connections results in a significant imbalance, increasing \incomefairness from 0.25 (0.26) to 0.39 (0.37) with $\mp=16$ ($\mp=30$). Using the pairwise limited free service instead reduces the imbalance. 


    \section{IPFS dataset}
\label{app:ipfsdataset}
In order to conduct simulations utilizing the proposed model, we have employed a publicly available dataset from IPFS, provided by the Protocol Labs~\cite{IPFS_gateway_data_paper}. This dataset~\footnote{bafybeiftyvcar3vh7zua3xakxkb2h5ppo4giu5f3rkpsqgcfh7n7axxnsa}, sourced from an official gateway, collected over $24$ hours, covering $7.1$M user requests, comprises various columns of data, including requested Content Identifier (CID) values by users and the corresponding bytes returned in response by the gateway. In our analysis, we assume that all CIDs exclusively pertain to entire files, with redundant CIDs indicating popular files.

\begin{figure*}[t]
    \centering
    \begin{subfigure}[b]{0.45\textwidth}
        \includegraphics[width=0.9\textwidth]{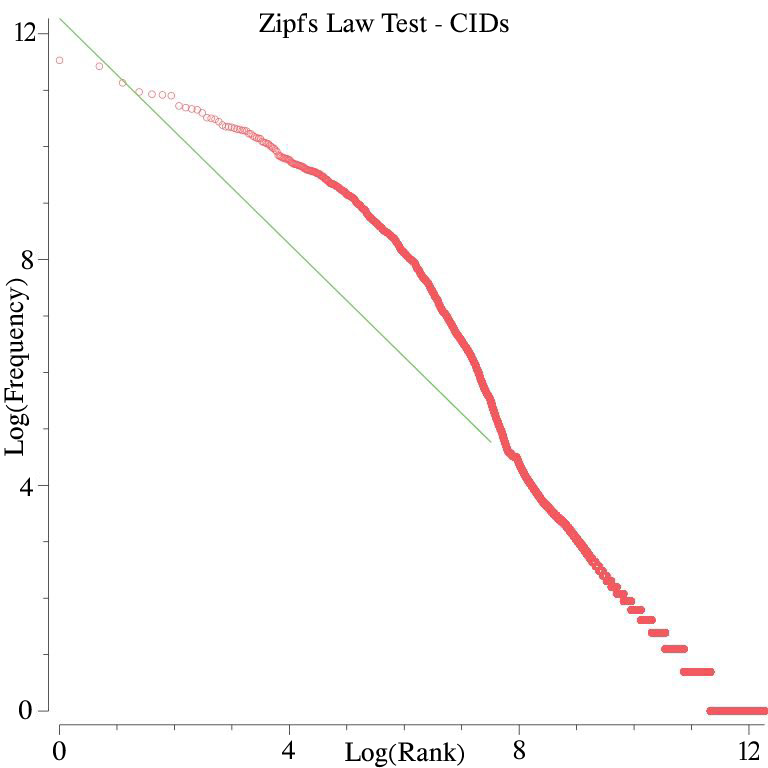}
        \caption{Distribution of CIDs in IPFS dataset}
        \label{subfig:cid-zipf}
    \end{subfigure}
    \hfill
    \begin{subfigure}[b]{0.45\textwidth}
        \includegraphics[width=0.9\textwidth]{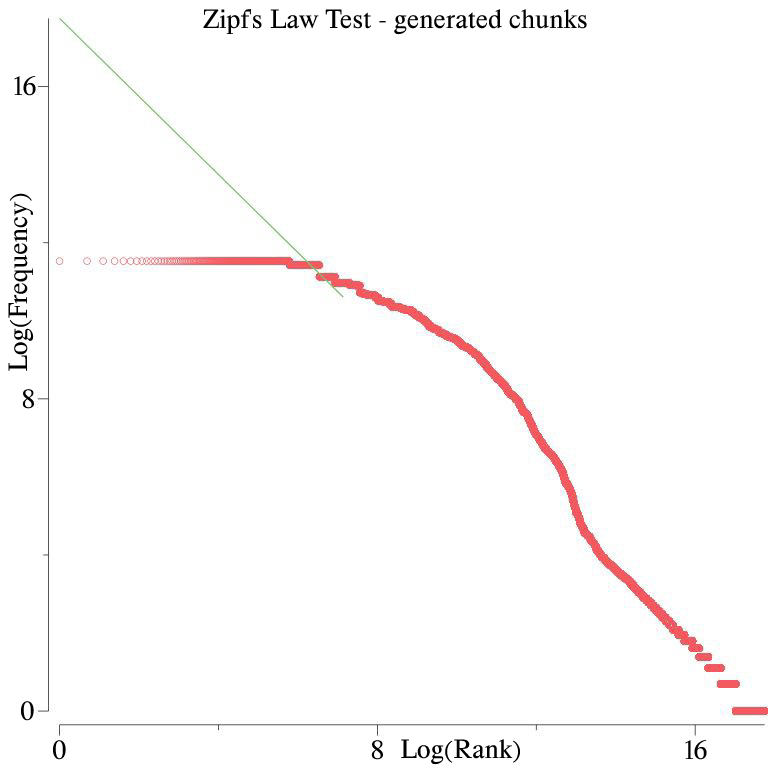}
        \caption{Distribution of randomly generated chunks using the CIDs}
        \label{subfig:chunks-zipf}
    \end{subfigure}
    \caption{Dataset distribution test}
    \label{fig:zipf-dist}
\end{figure*}

Within the context of our simulation, we extract vital information from this dataset, primarily focusing on the CIDs themselves, their respective frequencies, and the volume of bytes returned. Given that our simulation operates at the chunk level rather than dealing directly with entire files, we need to generate these chunks from the provided CIDs to facilitate our experiments. To achieve this, we employ the information regarding bytes returned, enabling us to generate random chunks while ensuring that all chunks derived from the same CID are identical in content. To visually represent the data distribution and the generated chunks stemming from it, we present Figure~\ref{fig:zipf-dist}. This figure illustrates the distribution of CIDs and the associated randomly generated chunks. It is important to note that the distribution closely adheres to the Zipf distribution, a characteristic confirmed through both graphical plots and statistical tests. 

\begin{figure*}[t]
    \begin{subfigure}[t]{0.45\textwidth}
        \centering
        \resizebox{\textwidth}{!}{%
            \begin{tikzpicture}
\pgfplotsset{width=7.5cm,height=6cm,compat=1.9}
\def\datafileA{data/baseincomefairnessT10sk8Th0Fg0W16.csv}
\pgfplotstableread{\datafileA}\datatable
\def\datafileB{data/baseincomefairnessT10sk8Th0Fg0W30.csv}
\pgfplotstableread{\datafileB}\datatable
\def\datafileC{data/baseincomefairnessT10sk16Th0Fg0W16.csv}
\pgfplotstableread{\datafileC}\datatable
\def\datafileD{data/baseincomefairnessT10sk16Th0Fg0W30.csv}
\pgfplotstableread{\datafileD}\datatable

\begin{axis}[
    tick label style={font=\small},
    legend style={anchor=north},
    xlabel={$\%$ of originators},
    ymin=0,
    ymax=1,
    ylabel={\incomefairness},
    xtick={1,10,50,100},
    legend pos=north east,
    ymajorgrids=true,
    legend entries={rand. k=8 $\omega=16$, rand. k=8 $\omega=30$, rand. k=16 $\omega=16$, 2choices k=8 $\omega=16$, 2choices k=8 $\omega=30$},
]
\addplot [mark=*, mark options={solid}, smooth, purple!80!white] table [x=originators, y=avg, col sep=comma]{\datafileA};
\addplot [mark=square*, mark options={solid}, smooth, blue!60!white] table [x=originators, y=avg, col sep=comma]{\datafileB};
\addplot [mark=x, black]table [x=originators, y=avg, col sep=comma]{\datafileC};
\addplot [mark=o, mark options={solid}, dashed, purple!80!white] table [x=originators, y=second, col sep=comma]{\datafileA};
\addplot [mark=square, mark options={solid}, dashed, blue!60!white] table [x=originators, y=second, col sep=comma]{\datafileB};
\end{axis}
\end{tikzpicture}
        }
        \caption{Result of selected experiments using the \textbf{generated} data.}
        \label{subfig:incomeFairnessXgatewaysGeneratedData}
    \end{subfigure}
    \begin{subfigure}[t]{0.45\textwidth}
        \centering
        \resizebox{\textwidth}{!}{%
            \begin{tikzpicture}
    \pgfplotsset{width=7.5cm,height=6cm,compat=1.9}
    \def\datafileA{data/IPFS_baseincomefairnessT10sk8Th0Fg0W16.csv}
    \pgfplotstableread{\datafileA}\datatable
    \def\datafileB{data/IPFS_baseincomefairnessT10sk8Th0Fg0W30.csv}
    \pgfplotstableread{\datafileB}\datatable
    \def\datafileC{data/IPFS_baseincomefairnessT10sk16Th0Fg0W16.csv}
    \pgfplotstableread{\datafileC}\datatable
    
    \begin{axis}[
        tick label style={font=\small},
        legend style={anchor=north},
        xlabel={$\%$ of gateways},
        ymin=0,
        ymax=1,
        ylabel={\incomefairness},
        xtick={1,10,50,100},
        legend pos=north east,
        ymajorgrids=true,
        legend entries={rand. k=8 $\omega=16$, rand. k=8 $\omega=30$, rand. k=16 $\omega=16$, 2choices k=8 $\omega=16$, 2choices k=8 $\omega=30$},
    ]
    \addplot [mark=*, mark options={solid}, smooth, purple!80!white] table [x=originators, y=avg, col sep=comma]{\datafileA};
    \addplot [mark=square*, mark options={solid}, smooth, blue!60!white] table [x=originators, y=avg, col sep=comma]{\datafileB};
    \addplot [mark=x, black]table [x=originators, y=avg, col sep=comma]{\datafileC};
    \addplot [mark=o, mark options={solid}, dashed, purple!80!white] table [x=originators, y=second, col sep=comma]{\datafileA};
    \addplot [mark=square, mark options={solid}, dashed, blue!60!white] table [x=originators, y=second, col sep=comma]{\datafileB};
    \end{axis}
    \end{tikzpicture}
        }
        \caption{Result of selected experiments using the \textbf{IPFS} data.}
        \label{subfig:incomeFairnessXgatewaysIPFSData}
    \end{subfigure}
    \caption{Comparison between the result with generated data vs. IPFS data.}
    \label{fig:ipfs_comapre_generated}
\end{figure*}
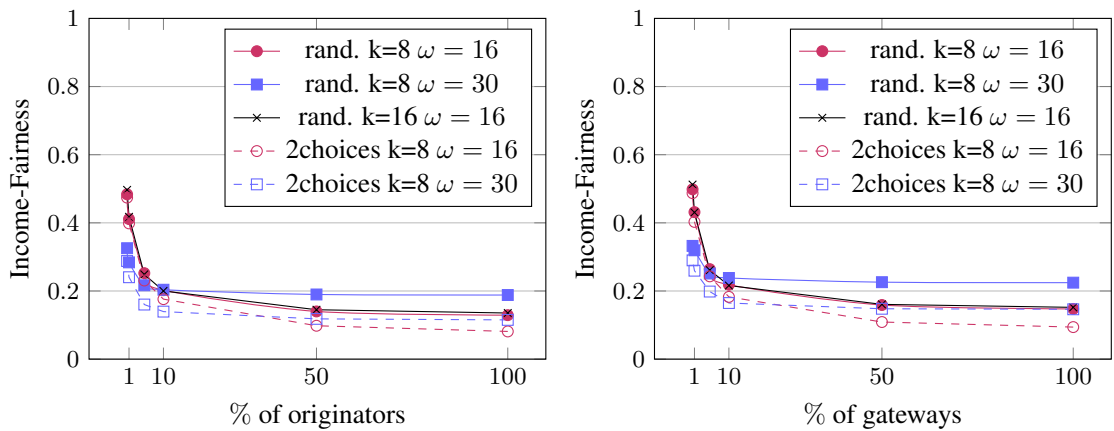

Figure~\ref{fig:ipfs_comapre_generated} shows the results of selected experiments conducting the generated and IPFS data. It attests that the results are the same and proves that our simulation will generate the same results if it uses the real workload. 

    \section{Caching chunks}
\label{app:caching}

One of the functionalities within a decentralized p2p network is the capability to cache data, which serves to enhance redundancy while concurrently reducing download latency.
The benefit of being able to cache chunks can be looked at from two perspectives. The first one is at the level of node operators, where for retrieving a cached chunk, they do not need to use their accounting units and bandwidth for forwarding the request. 
The second perspective is on the system level, where cached chunks mean redundancy, and fewer hops for downloading a chunk means less latency.

In Forwarding Kademlia, chunks are returned on the same path as the initial request, allowing the chunk to be cached at forwarding peers.


Swarm uses the least recently used (LRU) approach, where the least recently used chunk is evicted from the cache, where necessary.

%
Within the scope of our simulation, we have used two distinct eviction strategies, LRU, and Least Frequently Used (LFU).
The LRU strategy closely mirrors Swarm's caching model, emphasizing the removal of the least recently used chunks when eviction actions are triggered. The LFU approach adopts a different procedure, prioritizing the retention of popular chunks within the cache while evicting less popular ones, particularly if the cache approaches its predefined capacity limit. We use mainly LRU, but also LFU in some cases, where we found it to be the more effective strategy.


Table~\ref{tab:variable_cache_size} shows the result of experiments with $20\%$ of peers as originators with limited free service using standard and pairwise parameters and different cache sizes, where all peers use the LRU policy when eviction is needed. The \incomefairness among forwarding peers with cache size $5,000$ reaches the same fairness with the experiment without caching. Here, we only look at the fairness metric among the forwarding peers because originators have negative incomes that make the \incomefairness close to 1.

\begin{figure}
    \centering
\begin{tikzpicture}
    \pgfplotsset{width=6cm, height=4.5cm,compat=1.9}
    \def\datafileA{data/O20VaryingCacheSize.csv}
    \pgfplotstableread{\datafileA}\datatable
    \begin{axis}[
        tick label style={font=\small},
        xlabel={Cache size (\# of chunks)},
        ymin=0,
        ymax=100,
        ylabel={Cache Hit Ratio},
        xtick={1000,2000,5000},
        legend style={at={(1.02, 0.5)},
	    anchor=west,legend columns=1, legend cell align=left},
        ymajorgrids=true,
        legend entries={all, gateways, adaptive, non-adaptive},
    ]
    \addlegendimage{mark=*, color=purple!80!gray}
    \addlegendimage{mark=*, color=blue!60!black}
    \addlegendimage{style=solid}
    \addlegendimage{style=dashed}
    
    \addplot [mark=*, mark options={solid}, smooth, purple!80!gray] table [x=size, y=cacheHitRatio, col sep=comma]{\datafileA};
    \addplot [mark=*, mark options={solid}, smooth, blue!60!black] table [x=size, y=GAbs, col sep=comma]{\datafileA};
    \addplot [mark=o, mark options={solid}, dashed, purple!80!gray] table [x=size, y=cacheHitRatioAdj, col sep=comma]{\datafileA};
    \addplot [mark=o, mark options={solid}, dashed, blue!60!black] table [x=size, y=GAbsAdj, col sep=comma]{\datafileA};
    \end{axis}
\end{tikzpicture}
    \caption{Cache hit ratio at all nodes and gateways, with limited free service, and adapted free service.}
    \label{fig:cache_hit_ratio}
\end{figure}
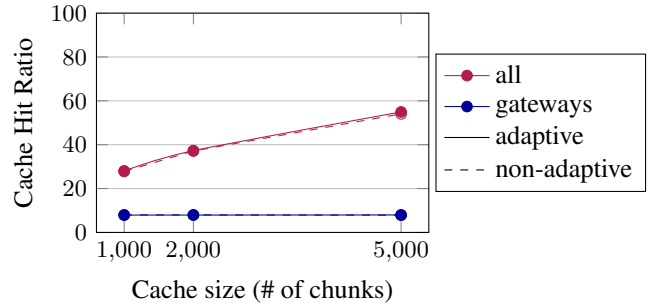

Figure~\ref{fig:cache_hit_ratio} also presents the cache hit ratio for all peers and originators, with an increasing trend for all peers, but an almost constant ratio for originators.

Now, we only talk about experiments with $50$ originators. 
We have evaluated the effect of different cache sizes and eviction policies for originators and non-originators and the result is reported in Table~\ref{tab:variable_cache_eviction_size} (the first two rows use the same cache size and eviction policy for both originators and forwarders). In this experiment, the default cache size for non-originator peers is $10,000$ chunks, and the eviction policy is LRU, while we vary the cache size for originators, and also the eviction policy to LFU (increasing cache hit-ratio). This experiment shows that using cache size $100,000$ for originators with LFU in the pairwise free service setting goes toward the level of fairness as the adaptive free service without caching but never reaches the same fairness. 

In conclusion, caching can affect fairness, especially with few originators, making more imbalance workload distribution and, consequently imbalanced reward distribution.

\begin{table}[t]
    \centering
    \caption{Variable cache size with $20\%$ originators, with and without adaptive free service.}
    \label{tab:variable_cache_size}
    \begin{tabular}{|c|c|c|c|c|}
        \hline
        \begin{tabular}[c]{@{}l@{}}Cache \\size \end{tabular} & \begin{tabular}[c]{@{}l@{}}NonOriginators \\ \incomefairness \end{tabular} & \begin{tabular}[c]{@{}l@{}l@{}}Cache \\Hit\\ Ratio \end{tabular} & \begin{tabular}[c]{@{}l@{}l@{}}Originators \\Cache \\Hit Ratio\end{tabular} & \begin{tabular}[c]{@{}l@{}l@{}}Adapted \\free \\service\end{tabular} \\ \hline
        1000 & 0.263488 & 28.04 & 7.93 & No \\ 
        1000 & 0.240270 & 27.81 & 7.96 & Yes \\
        2000 & 0.259398 & 37.28 & 7.92 & No \\ 
        2000 & 0.231437 & 37.13 & 7.99 & Yes \\
        5000 & 0.268428 & 54.94 & 7.98 & No \\ 
        5000 & 0.240368 & 54.06 & 7.89 & Yes \\
        - & 0.271520 & - & - & No \\
        - & 0.255149 & - & - & Yes \\
        \hline
    \end{tabular}
\end{table}


\begin{table}[t]
    \centering
    \caption{Variable cache eviction policy and cache size.}
    \label{tab:variable_cache_eviction_size}
    \begin{tabular}{|c|c|c|c|c|}
        \hline
        \begin{tabular}[c]{@{}l@{}l@{}}Originator\\ Cache\\ Size\end{tabular} & \begin{tabular}[c]{@{}l@{}}NonOriginator\\ \incomefairness \end{tabular} & \begin{tabular}[c]{@{}l@{}l@{}}Cache\\Hit \\Ratio\end{tabular} & \begin{tabular}[c]{@{}l@{}l@{}}Originators\\ Cache \\Hit Ratio \end{tabular} & \begin{tabular}[c]{@{}l@{}l@{}}Adapted\\ free\\ service \end{tabular} \\ 
        \hline
        $10000^*$ & 0.734202 & 69.78 & 9.1646  & No\\ 
        $10000^*$ & 0.710534 & 70.45 & 9.1722  & Yes \\ 
        10000 & 0.733877 & 72.44 & 20.9496 & No \\ 
        10000 & 0.710594 & 72.84 & 20.8760 & Yes \\ 
        20000 & 0.728847 & 74.12 & 30.7570 & No \\ 
        20000 & 0.702531 & 74.62 & 30.9453 & Yes \\ 
        100000 & 0.712670 & 79.59 & 61.3787 & No \\ 
        100000 & 0.666139 & 79.99 & 61.2868 & Yes \\
        - & 0.598561 & - & - & No \\
        - & 0.567436 & - & - & Yes \\ \hline
    \end{tabular}
\end{table}

    \section{Shuffling}
\label{app:shuffling}

In the paper, we showed that shuffling can help with the \incomefairness. In this section, we discuss the challenges using shuffling might have.

\subsection{Implemeneting Neighbor Shuffling}
To bootstrap into the network, nodes ping a few bootstrap
nodes, who broadcast their ID and new nodes can connect to
them. A similar mechanism can be used to find new nodes for
the creation of new or shuffled connections. To do so, a node
can ping a number of its immediate neighbors to broadcast
their willingness to find new neighbors. Nodes that have free
slots in their buckets can then offer connection.

In a saturated network, however, every node’s buckets are
filled and the node is not willing to connect to a new node.
In such situations, not only does the shuffling mechanism
perform poorly, but the network is less welcoming to new
nodes.

\subsection{Enforcing Shuffling}
Nodes already agree on the usage of a smart contract to settle
their payments. This smart contract can be further exploited to
also provide incentives to participate in the shuffling process.
Since this smart contract has access to the nodes’ balance, it
can be used to slash the nodes that do not shuffle frequently.
This can be determined by the amount of money transferred
from the same node ID in case of node ID shuffling, and the
amount of pairwise transfers in case of neighbors shuffling.

    \section{Payment models}
    \label{app:paymentmodels}
    \lj{Need to proof read this appendix.}

    When reaching the limit of the debt threshold, peers need to decide to either wait for the \refreshrate, and to rely on the limited free service or to settle the debt by transferring tokens. 
    Below, we investigate different strategies of how peers react once they cannot retrieve a chunk without settlement.
    We assume that all peers make use of reciprocity and do not settle debt unless it is required. 
    Peers, therefore, try to request chunks from a neighbor, from which the pairwise balance allows retrieval without a settlement.

    We investigate two different settlement strategies when a balance reaches the threshold.
    In the first strategy, peers only pay for the \textbf{C}urrent request. 
    Thus, when a peer wants to send a request, which would cause a connection's balance to pass the threshold, the peer issues a transfer for the amount necessary to receive this \textbf{C}urrent request.
    In a second strategy, peers pay for the \textbf{F}ull amount of debt and current request together. 
    These strategies represent peers who either try to minimize the transferred amount or the number of transfers issued.
    
    We use three different settings for which peers are willing to perform settlements when they are unable to retrieve a chunk within their reciprocity threshold.\\
    In setting \textbf{N}, \textbf{N}o one performs settlements.
    This setting is useful to investigate what download rate a specific \refreshrate allows peers to receive for free.\\
    In setting \textbf{O}, \textbf{O}nly originators perform settlements. This setting is similar to the default implemented in the Swarm network, with the difference that we assume originators also make use of free bandwidth if available. While not implemented as default, the Swarm network does not prohibit or prevent this policy; therefore, we find it more realistic.\\
    In setting \textbf{A}, \textbf{A}ll peers perform settlement.
    
    With setting \textbf{N}, but also \textbf{O}, it can happen that a given chunk cannot be retrieved.
    This happens if a peer on the route is unwilling to settle debt, but has reached the debt threshold with all relevant connections.
    We again implement policies on how originators react in this case.
    First, originators can \textbf{W}ait for balances to be refreshed and only continue other downloads once the current chunk has been received. 
    Second, originators can \textbf{G}ive up on a chunk and continue with another chunk.
    Both options may be somewhat unrealistic since a real client might neither abandon a needed chunk nor cease all other requests while waiting for this chunk.
    However, when evaluating the achieved download rate, the two policies, \textbf{W} and \textbf{G}, can be seen as best and worst-case scenarios for what may be achieved with more realistic measurements.
    
    In variant \textbf{A}, where all peers are willing to settle, originators do not need to give-up, or wait.
    However, this variant has a different problem, namely \emph{negative income}.
    If forwarding peers perform settlement, situations may arise where a forwarding peer transfers more utility tokens than he receives. 
    

    \subsection{Payment model summary}
    \label{subsect:model_summary}
    
    Table~\ref{tab:payment_models} gives a summary of the combination of different payment models we investigate, together with 3 letter acronyms.
    The different settings describe which peers perform settlements (\textbf{N},\textbf{O}, and \textbf{A}), how threshold rejections are handled (\textbf{G} and \textbf{W}), and
    how settlements are performed (\textbf{C} and \textbf{F}).
    We use \_ to replace a letter, where a given setting is not applicable. 
    For example, when no peers perform settlements (\textbf{N}), settlement policies (\textbf{C} and \textbf{F}) are not applicable. 
    Similarly, if all peers perform settlements (\textbf{A}), the handling of threshold rejections is not applicable (\textbf{G} and \textbf{W}).
    
    \begin{table}[]
        \centering
        \caption{Payment models acronyms and description}
        \label{tab:payment_models}
        \begin{tabular}{|l|l|}
            \hline
            Acronym  & Description   \\
            \hline
            \textbf{NG\_} & \begin{tabular}[c]{@{}l@{}}\textbf{N}o one performs settlements,\\ and originators \textbf{G}ive up requests exceeding the threshold.\end{tabular}\\ \hline
            \textbf{NW\_} & \begin{tabular}[c]{@{}l@{}}\textbf{N}o one performs settlements,\\ and originators \textbf{W}ait for the \refreshrate.\end{tabular}\\ \hline
            \textbf{OGF} & \begin{tabular}[c]{@{}l@{}}\textbf{O}nly originators perform settlements for the \\\textbf{F}ull amount, and \textbf{G}ive up on requests.\end{tabular}\\ \hline
            \textbf{A\_F} & \begin{tabular}[c]{@{}l@{}}\textbf{A}ll perform settlements for the \textbf{F}ull amount.\end{tabular}\\ \hline
            \textbf{A\_C} & \begin{tabular}[c]{@{}l@{}}\textbf{A}ll perform settlements for the \textbf{C}urrent request.\end{tabular}\\ \hline
            \textbf{OWF} & \begin{tabular}[c]{@{}l@{}}\textbf{O}nly originators perform settlements for the \\ \textbf{F}ull amount, and  \textbf{W}ait for the \refreshrate.\end{tabular}\\ 
            \hline
        \end{tabular}
    \end{table}

\subsection{Payment models impact}
\label{app:paymodelimpact}
This section studies the effect of reciprocity and limited free service under different payment models.
To provide limited free service, peers forgive a small fraction of debt every second.
As Subsection~\ref {subsec:recipAndLimitedFreeService} explains, the \refreshrate determines how many accounting units a peer is forgiven by its neighbors every second.
We conduct simulations to determine what download rates peers can get for free.

\begin{figure*}[t]
    \centering %
    \begin{subfigure}{0.45\textwidth}
        \centering
        \resizebox{\textwidth}{!}{
            \begin{tikzpicture}
            \pgfplotsset{width=7.5cm,compat=1.9}
            \def\datafileA{data/dl_rate2.dat}
            \pgfplotstableread{\datafileA}\datatable
            \begin{axis}[
            tick label style={font=\footnotesize},
            label style={font=\footnotesize},
            legend style={anchor=north,font=\scriptsize},
            xlabel={Requests per gateway (kB/s)},
            ylabel={Download rate (kB/s)},
            xmin=50, xmax=400,
            ymin=0, ymax=200,
            xtick={50, 100, 200, 400},
            ytick={50, 100, 150, 200},
            legend pos=north east,
            ymajorgrids=true,
            legend cell align={left},
            legend style={at={(0.5, 1.2)},
                   anchor=north,legend columns=2},
            cycle multi list={
                blue!60!black,  mark=*, mark options={solid}\\
                blue!60!black,dashed, mark=x, mark options={solid}\\
                blue!60!black,densely dashed, mark=o, mark options={solid}\\
                purple!80!gray,  mark=*\\
                purple!80!gray,dashed,  mark=x, mark options={solid}\\
                purple!80!gray,densely dashed, mark=o, mark options={solid}\\
            },
            legend entries={NG\_ 10\%, NG\_ 5\%, NG\_ 1\%, NW\_ 10\%, NW\_ 5\%, NW\_ 1\%}, 
            legend cell align={left}
            ]
            \addplot table [x=dl_rate, y=NG]{\datafileA};
            \addplot table [x=dl_rate, y=NG500]{\datafileA};
            \addplot table [x=dl_rate, y=NG100]{\datafileA};
            \addplot table [x=dl_rate, y=NW]{\datafileA};
            \addplot table [x=dl_rate, y=NW500]{\datafileA};
            \addplot table [x=dl_rate, y=NW100]{\datafileA};
            \end{axis}
            \end{tikzpicture}
        }
        \caption{Effective download rate for free service, when no peers perform settlements. 1\%, 5\%, and 10\% of peers act as gateways.}
        \label{fig:dl_rate_effective_free}
    
    \end{subfigure}
    \hfill
    \begin{subfigure}{0.45\textwidth}
        \centering
        \resizebox{\linewidth}{!}{
            \begin{tikzpicture}
                \pgfplotsset{width=7.5cm,compat=1.9}
                \def\datafileA{data/dl_rate.dat}
                \pgfplotstableread{\datafileA}\datatable
                \begin{axis}[
                    tick label style={font=\footnotesize},
                    label style={font=\footnotesize},
                    legend style={anchor=north,font=\scriptsize},
                    xlabel={Requests per gateway (kB/s)},
                    ylabel={Download rate (kB/s)},
                    xmin=50, xmax=400,
                    ymin=0, ymax=400,
                    xtick={50, 100, 200, 400},
                    ytick={100, 200, 300, 400},
                    legend pos=north east,
                    ymajorgrids=true,
                    legend cell align={left},
                    legend style={at={(0.5, 1.2)},
                           anchor=north,legend columns=2},
                    cycle multi list={
                        orange, mark=x, mark options={solid}\\
                        pink!90!gray,mark=o, mark options={solid}\\
                        magenta!80!gray,mark=diamond, mark options={solid}\\
                        blue!60!black,  mark=*, mark options={solid}\\
                        purple!80!gray,  mark=*\\
                    },
                    legend entries={A\_F, OGF, OWF, NG\_, NW\_},
                    legend cell align={left}
                ]
                \addplot table [x=dl_rate, y=all_pay_full]{\datafileA};
                \addplot table [x=dl_rate, y=only_o_full_g]{\datafileA};
                \addplot table [x=dl_rate, y=only_o_full_wait]{\datafileA};
                
                \addplot table [x=dl_rate, y=nop_now]{\datafileA};
                \addplot table [x=dl_rate, y=nop_wait]{\datafileA};
                \end{axis}
                \end{tikzpicture}
        }
        \caption{Effective download rate with different payment models.}
        \label{fig:dl_rate_effective}
    \end{subfigure}
    \hfill
    \caption{Effective download rate 
    with different request rate and payment models.}
    \label{fig:dl_rate}
\end{figure*}
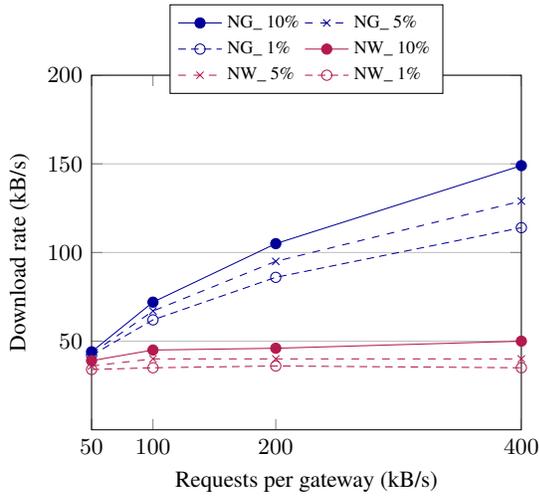
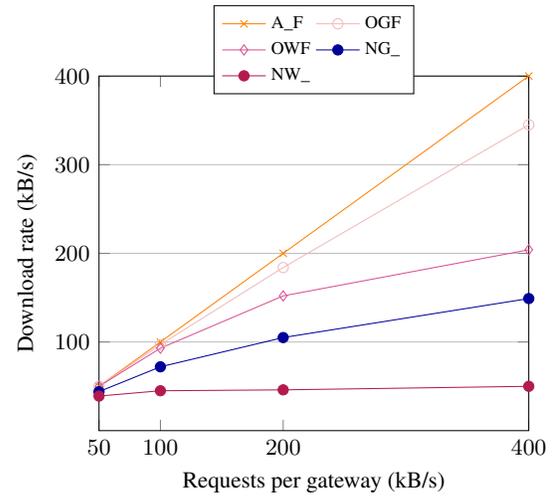

To measure the rate, we evaluate payment models where no peers perform settlements. 
Some requested chunks are not received in these models due to limits on the free service. 
Figure~\ref{fig:dl_rate_effective_free} shows the \textit{effective download rate}, calculated based on the ratio of successful downloads.

If originators retry a request, waiting for the~\refreshrate (NW\_), the effective download rate is almost constant.
If originators give up on requests exceeding the threshold (NG\_), they can achieve larger download rates for free. 
This is due to the uneven distribution of requests and accounting costs to buckets. 
For example, bucket 0 receives all requests for chunks located in the other half of the network. Also, these requests typically require to credit more accounting units.
Thus, while this variant allows downloading more chunks, different from the waiting variant, some requested chunks may never be downloaded.
We note that even with giving up, the download rate lies far below the request rate. 
Figure~\ref{fig:dl_rate_effective_free} also shows download rates for different numbers of originators. 
A larger number of originators allows a larger free download rate.

Figure~\ref{fig:dl_rate_effective} also shows download rates for other payment models.
In models where only originators perform settlements (OGF and OWF) download rates are still restricted by the limited free service.
In this model, forwarders need to rely on their free download rate.

However, if all peers perform settlements, another unfortunate behavior may occur.
Forwarders can end up with negative income when they pay for bandwidth.
Table~\ref{tab:negative-income} shows the models with peers with negative income at different checkpoints of the simulation. 
We see that our pairwise limited free service, paying for the full debt, and especially using the network generated by 2 choices reduces this problem.
We note that we ignore the transfers performed by originators for this evaluation.

\begin{table*}[t]
\caption{Percentage of the peers with negative income and sum of negative balances. The network generated with 2 choices solves the problem of negative income without changing the payment mechanism even in existence of greedy nodes that use their free slots instead of following Kademlia.}
\label{tab:negative-income}
\begin{tabular}{|l|l|l|l|l|l|l|l|}
\hline
\begin{tabular}[c]{@{}l@{}}Network\\ generation\end{tabular} & \begin{tabular}[c]{@{}l@{}}Payment\\ type \end{tabular}& \begin{tabular}[c]{@{}l@{}}
Greedy\\ use of\\ free\\ bandwidth
\end{tabular} & \begin{tabular}[c]{@{}l@{}}Pairwise\\ threshold \&\\ refresh rate\end{tabular} & \begin{tabular}[c]{@{}l@{}}Non-Originators \\ with negative \\ balance after \\ 10sec\end{tabular} & \begin{tabular}[c]{@{}l@{}}Sum\\ negative \\ balance\\ after\\ 10sec\end{tabular} & \begin{tabular}[c]{@{}l@{}}Non-Originators\\ with negative\\ balance after\\ 100sec\end{tabular} & \begin{tabular}[c]{@{}l@{}}Sum\\ negative\\ balance \\after\\ 100sec\end{tabular} \\ \hline
Random & per chunk & yes & no & 0.884\% & -4986 & 0.814\% & -43227 \\
Random & full debt & yes & no & 0.884\% & -5076 & 0.834\% & -47792 \\
Random & per chunk & no & no & 0.151\% & -530 & 0.121\% & -3239 \\
Random & full debt & no & no & 0.131 \% & -269 & 0.080\% & -2449 \\
Random & per chunk & yes & yes & 0.302\% & -473 & 0.101\% & -945 \\
Random & full debt & yes & yes & 0.181\% & -594 & 0.080\% & -2514 \\
Random & full debt & no & yes & 0.0\% & 0 & 0.0\% & 0 \\
2 choices & per chunk & yes & no & 0.0\% & 0 & 0.02\% & -13 \\ \hline
\end{tabular}
\end{table*}

%


\section{Discussion: Socialism or Free Enterprise?}
\label{app:Hardin}

In Hardin's reflections about the tragedy of commons~\cite{hardin1998extensions}, he describes a managed commons either by socialism or by the privatism of free enterprise: ``Either one may work; either one may fail: ‘The devil is in the details.’''
This section does not pretend to analyze both approaches; in the context of this paper, it suffices to say that both receive criticism and bring undesirable forms of centralization. 
Common criticisms of socialism are economic inefficiencies and the lack of individual freedom.
Free enterprise or capitalism raises concerns about income inequalities, monopolies, and oligopolies, which can be quickly summarized in the question: ``Do the rich get richer?''
We think, however, that some socialism and free enterprise thinking can reduce the \tft{} deficiencies in the social and economic spectrum. 

With respect to the economy, the pertinent questions are of the following types: 
What if we think about reciprocity beyond the closed incentivized system? 
A peer operator may participate in multiple systems simultaneously, for example, combining storage and financial networks, or in two distinct storage networks. 
If resources are scarce, how does the operator decide which is the most profitable network at a given time to share its resources?
It could also be the case that in one system, the peer is profitable and generates wealth, while in the other, it consumes and pays for resources in another.
Monetary or credit-based incentives can address the questions above by enabling more practical transactions among peers than in bartering economies. 
This area has generated significant interest~\cite{sirivianos2009robust,ghosh2014torpath,kopp2016koppercoin, miller2014permacoin}, especially with recent research focusing on cryptocurrencies and token-based incentives.  

On the other hand, there is the question of whether incentives align with the societal issues decentralized networks seek to address. 
How do humans operating a peer reciprocate the other humans operating peers, which may be doing good for the ecosystem in which the network is immersed? 
By doing good, we mean the wide spectrum of human activities, and it includes field workers such as investigative journalists and humanitarian aid working in remote disconnected villages or with restricted connectivity and the opposition activity in authoritarian regimes. 
Unfortunately, the economic model of Web 2.0 can expose vulnerable groups and facilitate surveillance from abusive governments~\cite{rod2015empowering}. 
For example, blogs published on centralized free blog servers may put the activist at risk. 
Will the incentives embedded in Web3 applications offer an alternative to these bloggers? 
Incentives alignment with societal issues is orthogonal with solutions that narrowly focus on eliminating free-riding but ignore the societal cornerstone aspect.

    \section{Which parameters make the difference on \incomefairness?}
\label{sec:workload_table}
To determine the optimal workload for use in our simulations and identify the key parameters impacting \incomefairness, 
we present the findings in Table~\ref{tab:different_exp_output}. The table reveals that workload size does not significantly influence changes in \incomefairness. 
However, variables such as the number of gateways, network size, and bucket size exhibit notable effects on \incomefairness. 
Therefore, we employ a reasonably large workload for our simulations.

\begin{table*}[t]
    \caption{The same experiment uses variable parameters (network size(n), bucket size(k), workload(million chunks)) to show their effect on \incomefairness and execution time.}
    \label{tab:different_exp_output}
    \centering
    \begin{tabular}{|c|c|c|c|c|c|c|}
    \hline
        \# originators & n & k & million chunks & \incomefairness & \begin{tabular}[c]{@{}l@{}}Non-originators\\ \incomefairness \end{tabular}& time(seconds) \\
    \hline
        50  & 1000  & 1  & 10   & 0.3663893333 & 0.344065     & 64 \\ 
        100 & 1000  & 1  & 10   & 0.3258393333 & 0.3182896667 & 63 \\ 
        50  & 1000  & 4  & 10   & 0.2829956667 & 0.2826903333 & 64 \\ 
        50  & 1000  & 1  & 100  & 0.3820393333 & 0.353684     & 615 \\ 
        50  & 10000 & 10 & 1000 & 0.3833496667 & 0.3833803333 & 293 \\ 
        50  & 20000 & 20 & 10   & 0.4286486667 & 0.4286206667 & 621 \\ 
        50  & 50000 & 50 & 10   & 0.4946106667 & 0.4945883333 & 1830 \\
    \hline
    \end{tabular}
\end{table*}
\end{appendix}

\end{document}